%% file: elsarticle-template-num.tex
\documentclass[preprint,12pt]{elsarticle}

\usepackage{amssymb}
\usepackage{amsmath}
\usepackage{algorithm}
\usepackage{algpseudocode}
\usepackage{booktabs}
\usepackage{multirow}
\usepackage{amssymb}
\usepackage{adjustbox}
\usepackage{float}
\usepackage{stfloats}
\usepackage{dsfont}
\usepackage{xcolor}
\usepackage{array}
\newcolumntype{P}[1]{>{\centering\arraybackslash}p{#1}}
\usepackage{amsthm}

\journal{Nuclear Physics B}

\begin{document}

\begin{frontmatter}



\title{UP2D: Uncertainty-aware Progressive Pseudo-label Denoising for Source-Free
Domain Adaptive Medical Image Segmentation}


\author[1]{Quang-Khai Bui-Tran\fnref{equal}}
\author[0]{Thanh-Huy Nguyen\fnref{equal}}
\author[1]{Manh D. Ho}
\author[1]{Thinh B. Lam}
\author[2]{Vi Vu}
\author[1]{Hoang-Thien Nguyen}
\author[3]{Phat Huynh}
\author[4]{Ulas Bagci\fnref{corres}}

\fntext[equal]{These authors contributed equally to this work.}
\fntext[corres]{Corresponding Author: Ulas Bagci (ulas.bagci@northwestern.edu)}
\affiliation[0]{
organization={Carnegie Mellon University},
            city={Pittsburgh},
            postcode={15213}, 
            state={PA},
            country={USA}}

\affiliation[1]{
organization={PASSIO Lab, North Carolina A\&T State University},
            city={Greensboro},
            postcode={27411},
            state={NC},
            country={USA}}

\affiliation[2]{
organization={Ho Chi Minh University of Technology},
            city={Ho Chi Minh},
            postcode={70000}, 
            country={Vietnam}}

\affiliation[3]{
organization={Industrial and Systems Engineering Department ,North Carolina A\&T State University},
            city={Greensboro},
            postcode={27411},
            state={NC},
            country={USA}}

\affiliation[4]{
organization={Northwestern University},
            city={Chicago},
            postcode={60611}, 
            state = {IL},
            country={USA}}

\input{sec/0_abstract}
\begin{graphicalabstract}
\includegraphics[width=1\linewidth]{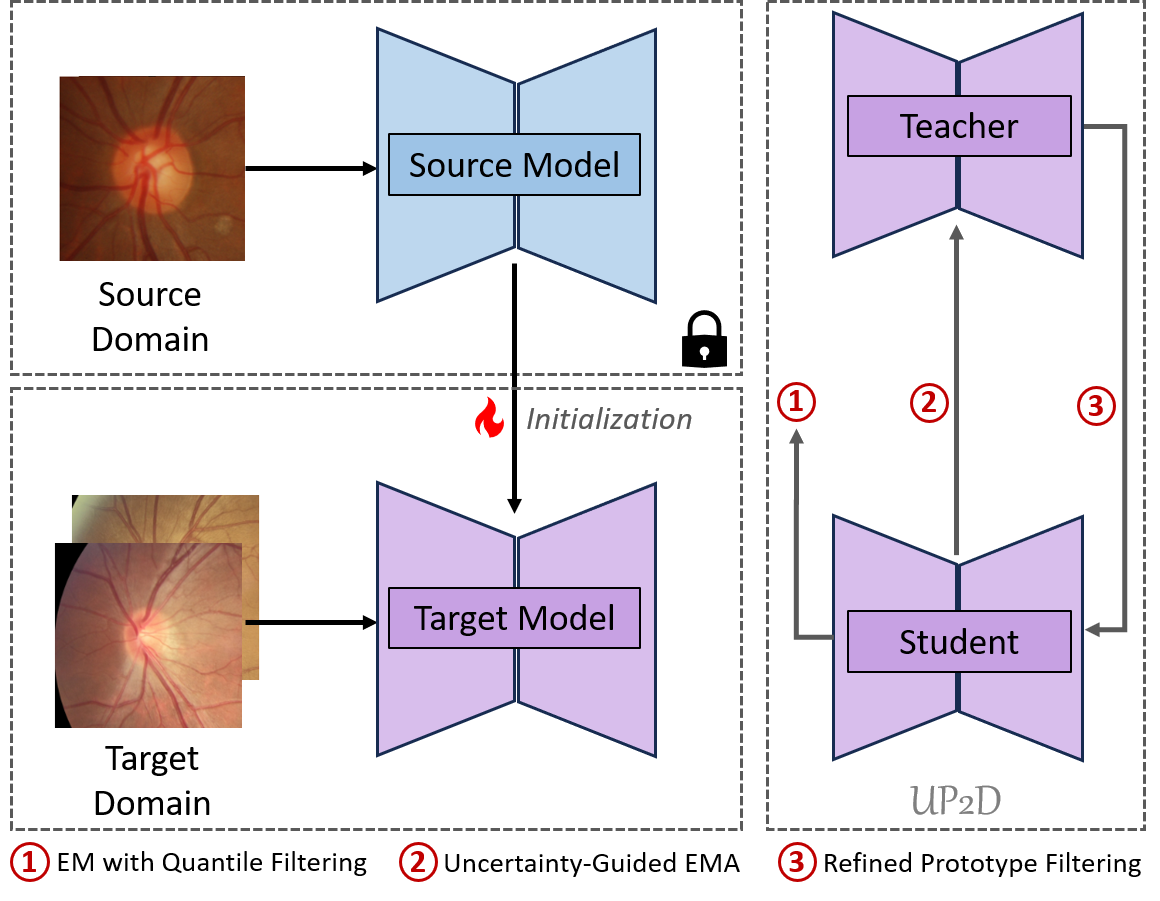}
\end{graphicalabstract}

\begin{highlights}
\item We propose an Uncertainty-aware Progressive Pseudo-label Denoising framework with a student–teacher pipeline in which the teacher generates pseudo-labels and denoises them via a \textit{Refined Prototype Filtering} mechanism. This design effectively leverages the progressively updated target-domain distribution, mitigates class imbalance by constructing low-uncertainty prototypes, and suppresses uninformative regions.
\item We design an Uncertainty-Guided EMA strategy that enables the teacher model to learn meaningful knowledge from the student while rejecting poor or unstable versions of the student model.
\item We introduce quantile-based filtering for entropy minimization, which  filters out high-confidence pixels based on the distribution of prediction probabilities to focus learning on uncertain regions.
\item Extensive experiments on two target domain fundus datasets and one open domain dataset demonstrate that our method achieves state-of-the-art performance both qualitatively and quantitatively while still having good generalization capability in the open domain.
\end{highlights}


\end{frontmatter}



\input{sec/1_intro}
\input{sec/related_works}
\input{sec/method}
\input{sec/exp}

\input{sec/conclusion}











\bibliographystyle{plain} 
\bibliography{references/ref_main, references/ref_sfda, references/ref_uda}
\end{document}

%% file: sec/0_abstract.tex
\begin{abstract}
Medical image segmentation models face severe performance drops under domain shifts, especially when data sharing constraints prevent access to source images. We present a novel Uncertainty-aware Progressive Pseudo-label Denoising (UP2D) framework for source-free domain
adaptation (SFDA), designed to mitigate noisy pseudo-labels and class imbalance during adaptation. UP2D integrates three key components: (i) a Refined Prototype Filtering module that suppresses uninformative regions and constructs reliable class prototypes to denoise pseudo-labels,
(ii) an Uncertainty-Guided EMA (UG-EMA) strategy that
selectively updates the teacher model based on spatially
weighted boundary uncertainty, and (iii) a quantile-based
entropy minimization scheme that focuses learning on ambiguous regions while avoiding overconfidence on easy pixels. This single-stage student–teacher framework progressively improves pseudo-label quality and reduces confirmation bias. Extensive experiments on three challenging retinal fundus benchmarks demonstrate that UP2D achieves state-of-the-art performance across both standard and open-domain settings, outperforming prior UDA and SFDA approaches while maintaining superior boundary precision.
\end{abstract}

%% file: sec/1_intro.tex
\section{Introduction}
\label{sec:introduction}
In recent years, medical image segmentation plays a pivotal role in a wide range of clinical tasks, including disease diagnosis, treatment planning, and surgical assistance \cite{multiresunet, v_net, Unet++}. Deep learning-based segmentation models, especially convolutional neural networks (CNNs) \cite{LeNet} and vision transformers \cite{ViT} have demonstrated impressive performance on various benchmarks, such as organ delineation in CT or lesion segmentation in retinal fundus images \cite{unet, UnetR}. However, the high accuracy of these models is often contingent upon the availability of large-scale, pixel-level annotated datasets, a requirement that is costly and labor-intensive, particularly in the medical domain where annotations must be provided by trained radiologists or specialists.

To alleviate the burden of annotation, Unsupervised Domain Adaptation (UDA) has been proposed to transfer knowledge from a labeled source domain to an unlabeled target domain. UDA techniques aim to bridge the distribution gap between domains by learning domain-invariant representations through adversarial learning, entropy minimization, or style transfer \cite{ADDA, DA_MaximumSquareLoss, FADA}.  In medical imaging, UDA is particularly important due to domain shifts caused by differences in imaging protocols, scanner vendors, or patient demographics \cite{kamnitsas2017unsupervised}. However, most UDA methods rely on access to both source and target data during training, which is often unrealistic in practice. Strict data privacy regulations, institutional policies, and patient consent limitations frequently prohibit the sharing of source data across medical centers.

In response to these constraints, Source-Free Domain Adaptation (SFDA) has recently emerged as a more practical and privacy-preserving alternative \cite{SHOT, sfda_fsm}. In the SFDA setting, only a trained source model is accessible, and adaptation must be conducted solely on unlabeled target data. This makes SFDA highly suitable for medical image analysis, where data-sharing restrictions are a major bottleneck. Despite its promises, SFDA introduces a new challenge: Without access to source data, model adaptation must relies on pseudo-labels generated by the source model, which can be noisy and unstable. Also, a more challenging problem, which is called source-free open-compound domain adaptation (SF-OCDA) \cite{zhao2022source}  is that when the model meets the open-domain datasets that are unseen before, it can drop marginally.

Although previous methods have achieved success in model adaptation, they still have some limitations. First, previous methods like CPR \cite{CPR}, and DPL \cite{DPL} are two-stage methods, explicitly separating the pseudo-label which is denoised through prototype filter process (created as fixed pseudo-labels by the frozen source model) and then training the adapted model of the noisy filtered pseudo label with a small number of epochs to avoid error accumulation during the adaptation phase. With this strategy, the predicted pseudo-label can be bias due to the source domain's distribution and also affect the denoising step, where it relies heavily on the prototype structured by the frozen source model because of this so the target model can not learn all the features and may not fully fit the target domain. Then the proposed student-teacher architecture, such as CBMT \cite{CBMT} runs a longer learning process but overlooks the unreliable of the teacher model which lacks any noisy label filtering during pseudo-label generation and suffers from error accumulation throughout the training process. Second, most models struggle to effectively address class imbalance, particularly for underrepresented classes such as the cup. During the adaptation phase, the model may degenerate due to the dominant class signal. This issue becomes more critical when deploying prototypes, as the prototype corresponding to the dominant classes suppresses the representation of minority ones. Finally is the uncertainty of boundary regions, due to the domain shift the segmentation model can be uncertain regions such as the boundary regions, previous works like PLPB \cite{PLPB} proposed to train the source with boundary loss and add the boundary pseudo label to the loss function for the adaptation process to focus more on the boundary but this apporach can not ensure that added boundary pseudo label always correct under heavy domain shift and also add computational complexity but also not all datasets provided the boundary ground truth.

To address the limitations of existing source-free domain adaptation (SFDA) approaches in medical image segmentation, we introduce a novel framework called Uncertainty-aware Progressive Pseudo-label Denoising (UP2D). Unlike prior two-stage methods that suffer from noisy supervision and limited adaptation, UP2D adopts a unified student–teacher architecture designed to iteratively refine pseudo-labels by leveraging the target-domain distribution while mitigating error accumulation. The teacher model, initialized from the source network, generates pseudo-labels using original target images and filters them through Refined Prototype Filtering (RPF) mechanism. This mechanism exploits the progressively updated target-domain distribution learned by the student and is further enhanced by uncertainty and region masking to suppress dominant classes' influence and improve the representation of underrepresented classes. To prevent confirmation bias, we introduce a selective update strategy, Uncertainty-Guided EMA (UG-EMA), that evaluates spatially weighted uncertainty and updates the teacher only when meaningful knowledge is distilled from the student. Furthermore, we incorporate an entropy minimization loss with quantile-based filtering, which restricts learning to low-confidence predictions and avoids over-penalizing already confident outputs, making it minimize evenly across whole images. Together, these components enable our method to produce reliable supervision and maintain robust performance even under severe domain shifts and open-domain settings.
The contributions of our work can be summarized as follows: 
\begin{itemize}
    \item We propose an Uncertainty-aware Progressive Pseudo-label Denoising framework with a student–teacher pipeline in which the teacher generates pseudo-labels and denoises them via a Refined Prototype Filtering mechanism. This design effectively leverages the progressively updated target-domain distribution, mitigates class imbalance by constructing low-uncertainty prototypes, and suppresses uninformative regions.
    \item We design an Uncertainty-Guided EMA strategy that enables the teacher model to learn meaningful knowledge from the student while rejecting poor or unstable versions of the student model.
    \item We introduce quantile-based filtering for entropy minimization, which effectively filters out high-confidence pixels based on the distribution of prediction probabilities to focus learning on uncertain regions.
    \item Extensive experiments on two target domain fundus datasets and one open domain dataset demonstrate that our method achieves state-of-the-art performance both qualitatively and quantitatively while still having good generalization capability in the open domain.
\end{itemize}

%% file: sec/related_works.tex
\section{Related Works}
\subsection{Unsupervised Domain Adaptation (UDA)}
Unsupervised domain adaptation (UDA) aims to adapt the knowledge learned in labeled source data to an unlabeled target domain. Common UDA approaches typically fall into two main categories: adversarial learning and pseudo-labeling strategies. 

Adversarial methods aim to bridge the domain gap by learning invariant features across domains using adversarial training schemes. For example, DANN \cite{DANN} uses a gradient reversal layer (GRL) \cite{GRL} to learn domain-invariant representations, ADDA \cite{ADDA} trains a separate target encoder with adversarial loss, CDAN \cite{CDAN} performs adversarial alignment conditioned on classifier's outputs to improve discriminability, DDA-Net \cite{DDA_net} applies dual-domain adaptation in both feature and image spaces for cross-modality segmentation. Pseudo-label-based methods seek to reduce domain shift by generating confident labels for the unlabeled target data. CBST \cite{CBST} generates class-balanced pseudo-labels for the target domain and iteratively retrains itself on these labels to improve domain alignment. ProDA \cite{ProDA} refines pseudo-labels by evaluating their distance to class prototypes, and aligns features to those prototypes during training. However, concerns regarding privacy and limitations in data transfer frequently render the traditional UDA setting infeasible in real-world applications.

\begin{figure*}[!ht]
    \centering
    \includegraphics[width=0.9\linewidth]{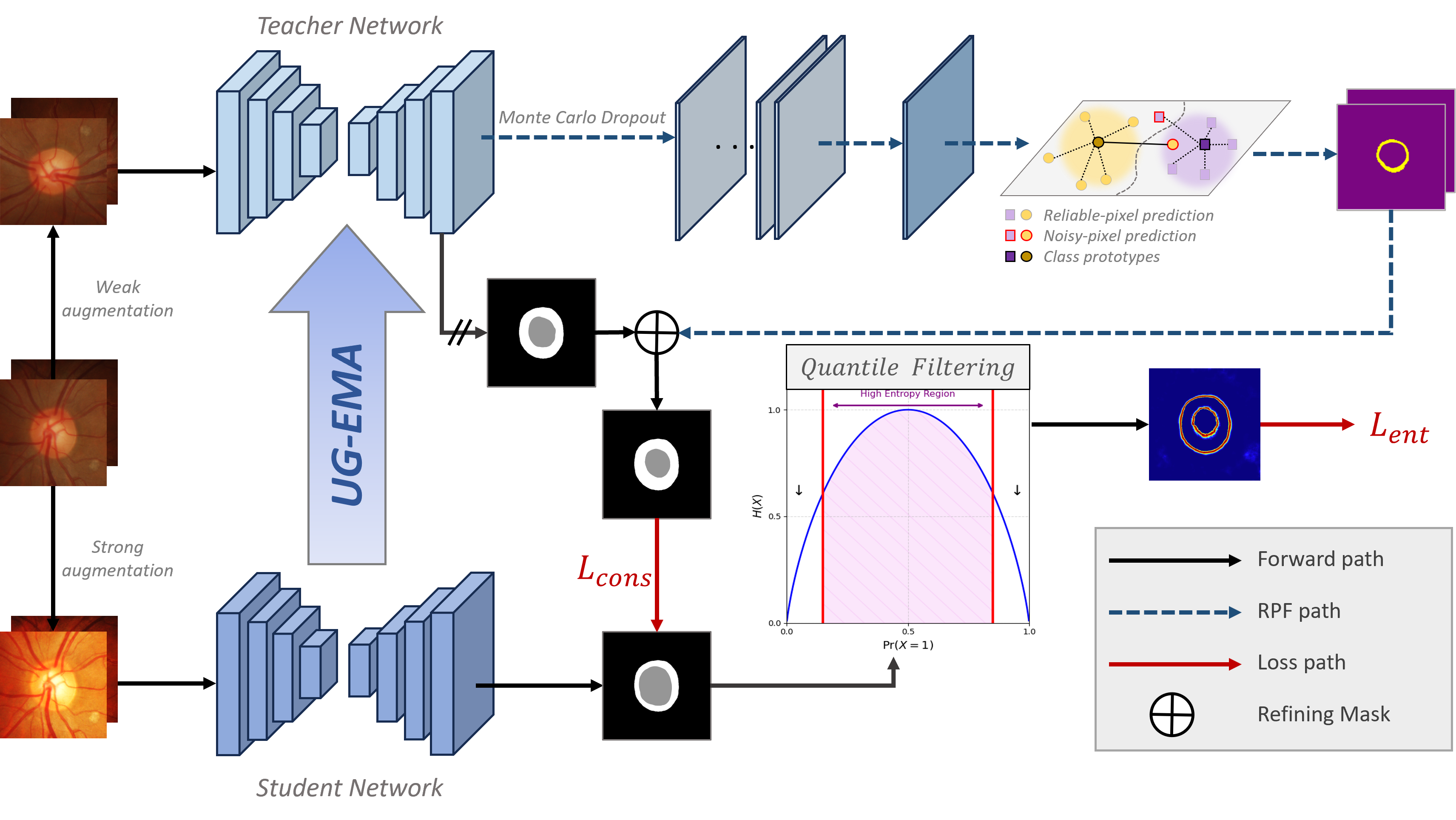}
    \caption{Overview of the proposed \textbf{UP2D} framework for source-free domain adaptive medical image segmentation. Our student-teacher framework enables the teacher to progressively generate new pseudo-labels that incorporate target-domain information and are denoised based on the updated model prototypes, which filter out noisy pixels for each pair of weakly and strongly augmented samples. Subsequently, a consistency loss is applied between the denoised pseudo-labels and the strongly augmented inputs of the student. Simultaneously, quantile-based entropy minimization is imposed on the student network to promote the reduction of uncertainty in ambiguous regions. Finally, the teacher is selectively updated through the \textbf{UG-EMA} decision-making strategy to exclude unreliable prediction versions.}
    \label{fig:main_fig}
\end{figure*}

\subsection{Source-Free Domain Adaptation (SFDA)}
Due to increasing concerns around data privacy, particularly in healthcare applications, patient data security is crucial. In addition, some data sources are too large and difficult to transfer. SFDA has emerged as a practical alternative to traditional UDA. Instead of requiring access to source data, SFDA methods rely solely on a pre-trained source model and adapt it to the target domain, typically through self-training frameworks.

Among early explorations, several methods adopt the mean-teacher paradigm. For instance, CBMT \cite{CBMT} introduces a two-stage teacher–student framework with a class-balanced loss that enhances performance on rare foreground classes in fundus segmentation. CrossMatch \cite{CrossMatch} extends this idea to cross-modal inputs, leveraging consistency between RGB and depth streams for robust segmentation.

Despite their success, teacher-based methods often suffer from confirmation bias when pseudo-labels are noisy, especially in dense prediction tasks like segmentation. To address this, another line of research focuses on pseudo-label denoising strategies. These methods aim to filter, correct, or reweight noisy predictions, thereby improving the quality of supervision and stabilizing adaptation. For example, DPL \cite{DPL} introduces a two-step denoising process that filters unreliable pseudo labels using both pixel-level uncertainty and class-level prototype similarity, ensuring that only confident predictions are used for training. Similarly, U-D4R \cite{U_D4R} adopts a coarse-to-fine denoising framework: it first selects labels based on adaptive class-wise thresholds, then refines them through an uncertainty-aware rectification mechanism. Another recent work, CPR \cite{CPR}, improves label quality by learning contextual similarity between pixels, enabling the model to revise and calibrate noisy pseudo labels using surrounding structure. 

%% file: sec/method.tex
\section{Methods}

Fig.~\ref{fig:main_fig} illustrates our proposed method. In this section, we introduce a unified single-stage framework inspired by the teacher ($f^T$) - student ($f^S$) paradigm. To mitigate error accumulation, we first design a Refined Prototype Filtering mechanism that selects reliable and informative pixels when transferring knowledge from the teacher to the student. Furthermore, we propose a decision-making UG-EMA strategy to evaluate aggregated uncertainty and decide whether to update the teacher model. Finally, we incorporate an entropy loss with quantile filtering to suppress high-entropy predictions in the target domain.

\subsection{Problem Formulation and Notation}
In the Source-Free Domain Adaptation (SFDA) setting, a labeled source dataset $\mathcal{D_{S}} = \{(x_s^i, y_s^i)\}_{i=1}^{N_S}$, is provided, where $y_s^i \in \{0,1\}^{H \times W \times C}$ is the ground truth segmentation mask and here $H$, $W$, and $C$ denote the height, width, and number of classes, respectively, with $C = 2$ since there are two segmentation targets: the optic cup and the optic disc. We first train a source model $f^{s}$ on $\mathcal{D_{S}}$ using cross-entropy loss. With the well-trained source model $f^{s}$, our final goal is to adapt it to obtain a target model $f^{t}$, using only an unlabeled target dataset $\mathcal{D_{T}} = \{x_t^i\}_{i=1}^{N_T}$.  

\subsection{Refined Prototype Filtering }
During adaptation, directly using noisy pseudo-labels from the teacher can lead to error accumulation in the student model. To mitigate this, inspired by \cite{snell2017prototypical}, we introduce a filtering mechanism to remove unreliable pixels from the teacher’s predictions. We estimate uncertainty using Monte Carlo Dropout \cite{gal2016dropout} with $K$ stochastic forward passes. For each pixel $v$, we compute predictions $p_{v,k} = f^T_v(x_t)$ for $k = 1, \ldots, K$, and derive the mean prediction $p_v = \text{avg}(p_{v,1}, \ldots, p_{v,K})$ and standard deviation map $u_v = \text{std}(p_{v,1}, \ldots, p_{v,K})$. Using a confidence threshold $\gamma$, the pseudo-label is then defined as $\hat{y}_v = \mathds{1}[p_v \geq \gamma]$. 

To build reliable prototypes, we retain only features and predictions corresponding to pixels with low uncertainty:
\begin{equation} \label{eq: unc_filt}
f = f \cdot \mathds{1}[u < \eta_1], \quad 
p = p \cdot \mathds{1}[u < \eta_1],
\end{equation} 
Where $f$ is the feature map extracted from the teacher's penultimate layer, $p$ is the prediction output of the teacher model for all pixels, $u$ is the standard deviation map for all pixels,  and $\eta_1$ is a predefined uncertainty threshold. For each class $\omega \in \{0,1\}$ (0 for background and 1 for foreground), we compute the class-specific prototype:
\begin{equation} \label{eq: prototype}
c^{\omega} = \frac{\sum_{v} f_v \cdot \mathds{1}[\hat{y}_v = \omega] \cdot p_v}{\sum_{v} \mathds{1}[\hat{y}_v = \omega] \cdot p_v}.
\end{equation}
Then, for each pixel, we calculate its distance to each prototype:
\begin{equation}\label{distance}
d_v^{\omega} = \left\lVert f_{v} - c^{\omega} \right\rVert_2.
\end{equation}
Based on these distances, we compute a denoising mask:
\begin{equation}
m_v =\mathds{1}[\hat{y}_v = 1] \cdot \mathds{1}[d_v^{1} < d_v^{0}]+ \mathds{1}[\hat{y}_v = 0] \cdot \mathds{1}[d_v^{1} > d_v^{0}].
\end{equation}
Finally, the consistency loss is masked with $m_v$ to focus learning on more reliable pixels:
\begin{equation}
\mathcal{L}_{\text{cons}} = -\sum_v m_v \Big[ \hat{y}_v \log(f_v^S(x_t)) + (1 - \hat{y}_v) \log(1 - f_v^S(x_t)) \Big],
\end{equation}
where $\hat{y}_v$ is the pseudo-label predicted at pixel $v$ from the teacher model, and $f_v^S(x_t)$ is the student model's output probability for pixel $v$.

In medical imaging, some classes often occupy only a small region and are often surrounded by dominant neighboring structures or background. Consequently, the standard feature computation in Eq.~\ref{eq: unc_filt} may place the background prototype too far from the object boundary or fail to separate noisy boundary regions due to the inclusion of uninformative, high-uncertainty pixels. This often results in incorrect filtering of boundary pixels, where misclassification is most likely. To address this, we first suppress high-confidence background pixels by masking those predicted as background by both the underrepresented class and its surrounding class classifiers. This preserves informative features for the small class of interest. Since this region may still contain noise, we further refine it by removing pixels with high uncertainty or entropy. The informative and uncertainty masks are defined as:

\begin{align}
    m^{w_1}_{info\_region} &= 1 - \mathds{1}[\hat{y}^{w_1} = \hat{y}^{w_2} = 0], \\
    m^{w_1}_{uncertainty} &= \mathds{1}[u < \eta_1] \cdot \mathds{1}[e < \eta_2],
\end{align}
where $w_1$ denotes the underrepresented class, and $w_2$ represents the surrounding outer class. Then, the feature map and predictions are computed as follows:
\begin{equation}
f^{w_1} = f^{w_1} \cdot m^{w_1}_{uncertainty} \cdot m^{w_1}_{info\_region},
\end{equation}
\begin{equation}
p^{w_1} = p ^{w_1}\cdot m^{w_1}_{uncertainty} \cdot m^{w_1}_{info\_region},
\end{equation}
where entropy map $e = - p \cdot \log(p)$ is computed from the output probability map $p$, $\eta_2$ is a predefined entropy threshold, and $\hat{y}_v^{w_1}$, $\hat{y}_v^{w_2}$ are predictions from the underrepresented class and surrounding outer class classifiers, respectively. After computing prototypes and pixel-wise distances using $f^{w_1}$ and $p^{w_1}$ as defined in Eq. \ref{eq: prototype} and  \ref{distance}, we define a denoising mask for the underrepresented class:
\begin{equation}
\label{eq:denoise_mask}
\begin{split}
m_v^{w_1} =\ & \mathds{1}[\hat{y}_v^{w_1} = 1] \cdot \mathds{1}[d_v^{1} < d_v^{0}] \\
& + \mathds{1}[\hat{y}_v^{w_1} = 0] \cdot 
 \mathds{1}[(\hat{y}_v^{w_2} = 0) \lor (d_v^{1} > d_v^{0})]. 
\end{split}
\end{equation}
This mask ensures that noisy pixels are correctly identified due to prototype refinement, focusing only on relevant features of the object boundaries, thereby improving representation in challenging scenarios.
\subsection{Uncertainty-Guided EMA}
To highlight the importance of boundaries in semantic segmentation, we introduce a Gaussian-based weighting mechanism that concentrates on transition regions between the foreground and background. Predictions near the object center are generally more confident, while boundaries are less certain and harder to predict. To address this, we apply a 2D Inverted-Gaussian weighting map that attains its minimum at the center and increases toward the edges, thereby emphasizing uncertain boundary regions. Let $(\mu_x^k, \mu_y^k)$ denote the center of the foreground region for class $k$, computed from its binary mask, while the standard deviations $\sigma_x^k$ and $\sigma_y^k$ are derived from the spatial distribution of that region:

\begin{align}
    \sigma_x^k &= s \cdot W_k, \quad \sigma_y^k = s\cdot H_k, \label{eq:sigma}
\end{align}
where $W_k$ and $H_k$ represent the width and height of the foreground region for class $k$, respectively, and $s$ is a predefined scaling factor. The 2D Inverted-Gaussian weight at pixel location $(x_v, y_v)$ is then defined as:
\begin{align}
\tilde{G}^k(x_v, y_v) = 1-\exp\left( -\frac{(x_v - \mu_x^k)^2}{2 (\sigma_x^k)^2} - \frac{(y_v - \mu_y^k)^2}{2 (\sigma_y^k)^2} \right).
\end{align}

Next, $\hat{y}_v$ is denoted as the binary pseudo-label from the teacher at pixel $v$,  $\delta$ is the ratio of the foreground region to the entire image, and \( \tilde{G}^k_v = \tilde{G}^k(x_v, y_v) \). To avoid including unrelated background regions, we define a Gaussian threshold $\tau$ that includes all the background pixels whose Gaussian weights lie below the maximum value within the foreground region (offset by $\delta$), while retaining all foreground pixels. The Gaussian threshold is defined as:
\begin{equation}
\tau = \max_v \{ \tilde{G}^k \mid \hat{y} = 1 \} - \delta,
\label{eq:threshold}
\end{equation}
and the binary mask $A_v$ at the pixel $v$ is given by:
\begin{equation}
A_v = \mathds{1}\Big[\hat{y}_v = 1 \lor \big(\hat{y}_v = 0 \land \tilde{G}^k_v \leq \tau \big)\Big],
\label{eq:mask}
\end{equation}

Finally, \( p^S_v \) is computed as the sigmoid output of the student model at the pixel \( v \). The current learning state of the student model is computed by the spatially weighted entropy estimation as follows: 
\begin{equation}
\mathcal{E} = - \frac{\sum_v A_v \cdot \tilde{G}^k_v \cdot p^S_v \log p^S_v}{\sum_v A_v \cdot \tilde{G}^k_v},
\label{eq:entropy_est}
\end{equation}

\begin{figure}[!ht]
    \centering
    \includegraphics[width=0.7\linewidth]{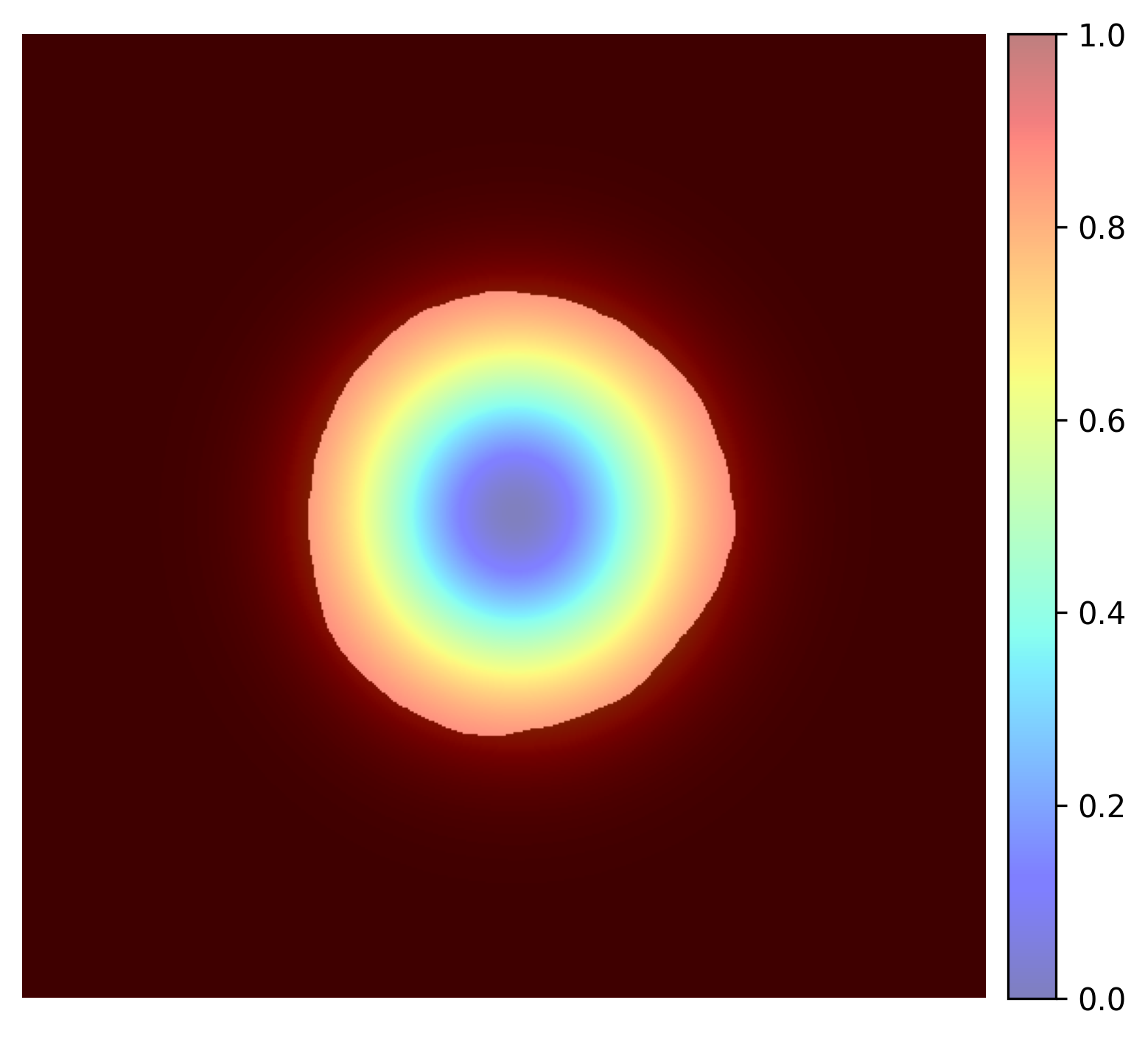}
    \caption{Inverted Gaussian map overlaid on the pseudo-label. The central regions are assigned low weights, while the edges receive higher weights, emphasizing boundary regions.}
    \label{fig:gauss_map}
\end{figure}

This formulation suppresses the influence of central regions while amplifying the contribution of spatially ambiguous boundary regions, as illustrated in Fig.~\ref{fig:gauss_map}. By leveraging updates guided by this formulation, the teacher model directs its learning attention toward more critical and uncertain areas, thereby enhancing boundary precision and strengthening overall segmentation robustness.

Previous EMA approaches in semi-supervised learning rely on supervised signals during training \cite{tarvainen2017mean}. In the source-free setting, where no labels are available, this must be carefully managed to prevent error accumulation. We introduce the Uncertainty-Guided EMA (UG-EMA) (Algorithm~\ref{alg:ewa-ema}), which updates the teacher model only when the student provides reliable feedback.

We track the lowest mean uncertainty $\bar{\mathcal{E}}^{\min}_e$ across epochs and initialize each new epoch's batch-level minimum uncertainty $\mathcal{E}^{\min}_b$ with this value. This avoids unstable updates caused by noisy per-batch minima $\mathcal{E}_b$, which may fluctuate due to varying batch composition. Each batch uncertainty is computed using Eq.~\ref{eq:entropy_est}. If a batch achieves $\mathcal{E}_b < \mathcal{E}^{\min}_b$, the teacher parameters $\theta_t$ are updated via EMA, and $\mathcal{E}^{\min}_b$ is set to $\mathcal{E}_b$. At the end of the epoch, the mean uncertainty $\bar{\mathcal{E}}_e$ is calculated and compared to $\bar{\mathcal{E}}^{\min}_e$, which is updated if a new minimum is found.

\begin{algorithm}
\begin{algorithmic}[1]
\State \textbf{Input:} Student $f^S(\cdot;\theta_s)$, Teacher $f^T(\cdot;\theta_t)$.
\State \textbf{Hyper-parameter:} Update rate $\alpha$, 
Epoch number $E$.
\State \textbf{Parameter:} Minimum epoch uncertainty $\bar{\mathcal{E}}^{\min}_e$; Minimum batch uncertainty within the current epoch $\mathcal{E}^{\min}_b$; Current batch uncertainty $\mathcal{E}_b$; Current epoch uncertainty $\bar{\mathcal{E}_e}$.
\State 
\State Initialize $\bar{\mathcal{E}}^{\min}_e \gets \infty$.
\For{$\text{epoch} = 1$ to $E$}
    \State $\mathcal{E}^{\min}_b \gets \bar{\mathcal{E}}^{\min}_e$
    \For{each batch in epoch}
        \State Compute $\mathcal{E}_b$ using Eq.~\ref{eq:entropy_est}.
        \If{$\mathcal{E}_b < \mathcal{E}^{\min}_b$}
            \State $\theta_t \gets \alpha \cdot \theta_t + (1 - \alpha) \cdot \theta_s$
            \State $\mathcal{E}^{\min}_b \gets \mathcal{E}_b$
        \EndIf
    \EndFor
    \State Compute $\bar{\mathcal{E}_e} \gets \frac{1}{B} \sum_{b=1}^{B} \mathcal{E}_b$.
    \If{$\bar{\mathcal{E}}_e < \bar{\mathcal{E}}^{\min}_e$}
        \State $\bar{\mathcal{E}}^{\min}_e \gets \bar{\mathcal{E}}_e$
    \EndIf
\EndFor
\caption{Uncertainty-Guided EMA}
\label{alg:ewa-ema}
\end{algorithmic}
\end{algorithm}

\subsection{Entropy Minimization with Quantile Filtering}
Existing SFDA methods \cite{SHOT} perform entropy minimization over the entire prediction map, which unintentionally includes already high-confident predictions with low entropy, where further minimization becomes redundant. To address this, we propose a quantile-based filtering strategy that adaptively focuses entropy minimization on uncertain predictions falling within a meaningful range.

Given a pixel-wise prediction from the student model, $p^S_v$, we first compute the lower and upper quantile thresholds $q_{\text{low}}$ and $q_{\text{high}}$ over the distribution of $P^S$ across the batch: 
\begin{align}
    &q_{\text{low}} = \operatorname{Quantile}(P^S, \beta), \\
&q_{\text{high}} = \operatorname{Quantile}(P^S, 1 - \beta),
\end{align}
where $\beta$ is a predefined quantile threshold, and $\operatorname{Quantile}(X, \beta)$ is the $\beta$-quantile function of the values in $X$ and defined by: 
\begin{align}
    \operatorname{Quantile}(X, \beta) = \sup\{x \in \mathbb{R} \mid F_X(x) \leq \beta \},
\end{align}
where $F_X(x)$ denotes the Cumulative Distribution Function of the random variable $X$.

\begin{table*}[!ht]
\centering
\caption{The quantitative results with different Source-Free methods on two datasets. \textbf{Bold} texts highlight the best scores.}
\setlength{\tabcolsep}{3mm}
\renewcommand{\arraystretch}{1}
\resizebox{1\textwidth}{!}{%
\begin{tabular}{|l|c|c|c|c|c|}
\hline
\multirow{2}{*}{\textbf{Method}} & \multirow{2}{*}{\textbf{S-F}} &
\multicolumn{2}{c|}{\textbf{Optic Disc Segmentation}} & 
\multicolumn{2}{c|}{\textbf{Optic Cup Segmentation}} \\
\cline{3-6}
& & Dice[\%] $\uparrow$ & ASSD[pixel] $\downarrow$ & Dice[\%] $\uparrow$ & ASSD[pixel] $\downarrow$ \\
\hline
\multicolumn{6}{|c|}{\textbf{Drishti-GS}} \\
\hline
Source only &  & $96.12\pm1.74$  & $4.40\pm1.93$ & $83.80\pm12.82$ & $10.69\pm6.95$ \\
Target only &  & $97.06\pm1.17$ & $3.25\pm1.20$ & $89.36\pm8.51$ & $7.07\pm3.50$ \\
\hline
BEAL (MICCAI'19) \cite{BEAL}  & $\times$ & $96.12\pm1.53$  & $4.48\pm1.84$ & \underline{$85.18\pm11.86$} & \underline{$9.66\pm6.28$}\\
AdvEnt (CVPR'19) \cite{AdvEnt} & $\times$ & $93.09\pm3.27$  & $8.55\pm5.32$ & $80.39\pm13.91$ & $13.01\pm6.66$ \\
TENT (ICLR'21) \cite{Tent} & \checkmark & $92.60\pm2.82$  & $8.93\pm4.05$ & $79.97\pm12.69$ & $13.39\pm7.06$ \\
DPL (MICCAI'21) \cite{DPL}  & \checkmark & $93.13\pm1.86$  & $8.01\pm1.97$ & $82.93\pm15.25$ & $11.61\pm7.59$ \\
CPR (MICCAI'23) \cite{CPR}  & \checkmark & $96.36\pm1.25$  & $4.09\pm1.33$ & $83.81\pm14.72$ & $10.96\pm7.02$ \\
CBMT (MICCAI'23) \cite{CBMT}   & \checkmark & $\underline{96.61\pm 1.45}$  & $\underline{3.85\pm1.63}$ & $84.33\pm11.70 $ & $10.30\pm5.88$ \\
PLPB (WACV'24) \cite{PLPB} & \checkmark & $93.82\pm2.04$  & $7.51\pm2.74$ & $84.36\pm10.59$ & $10.29\pm4.53$ \\
SBIF (ISBI'25) \cite{yaacovi2025source}  & \checkmark & $96.59\pm1.18$ & $3.92\pm1.29$ & $84.47\pm11.41$ & $10.21\pm5.79$ \\
\textbf{Ours}              & \checkmark & 
\textbf{96.61 $\pm$ 1.28} & \textbf{3.83 $\pm$ 1.42} & \textbf{86.61 $\pm$ 12.38} & \textbf{8.78 $\pm$ 5.45} \\
\hline
\multicolumn{6}{|c|}{\textbf{RIM-ONE-r3}} \\
\hline
Source only &  & $88.15\pm3.32$  & $11.15\pm3.30$ & $74.88\pm25.50$ & $7.87\pm4.45$ \\
Target only &  & $96.03\pm1.80$ & $3.42\pm1.50$ & $80.51\pm20.56$ & $6.83\pm5.57$ \\
\hline
BEAL (MICCAI'19) \cite{BEAL}   & $\times$ & $90.28\pm3.49$  & $8.95\pm3.23$ & $76.06\pm25.41$ & \underline{$7.19\pm3.91$} \\
AdvEnt (CVPR'19)  \cite{AdvEnt}  & $\times$ & $76.13\pm14.46$  & $23.30\pm12.81$ & $62.97\pm28.62$ & $11.58\pm5.28$ \\
TENT (ICLR'21)  \cite{Tent} & \checkmark & $82.33\pm9.08$  & $22.19\pm20.61$ & $78.01\pm16.32$ & $10.62\pm9.46$ \\
DPL (MICCAI'21) \cite{DPL} & \checkmark & $85.98\pm7.09$  & $18.23\pm7.16$ & $64.51\pm16.24$ & $15.26\pm11.40$ \\
CPR (MICCAI'23) \cite{CPR}  & \checkmark & $92.39\pm2.66$  & $6.86\pm2.35$ & $75.04\pm17.94$ & $10.43\pm5.05$ \\
CBMT (MICCAI'23) \cite{CBMT}  & \checkmark & $93.36\pm4.07$ & $6.20\pm4.79$ & $81.16\pm14.71$ & $8.37\pm6.99$ \\
PLPB (WACV'24) \cite{PLPB} & \checkmark & $83.75\pm5.68$  & $18.77\pm12.00$ & $73.39\pm18.50$ & $11.84\pm6.44$ \\
SBIF (ISBI'25) \cite{yaacovi2025source}  & \checkmark & \underline{$93.81\pm3.70$}  & \underline{$5.58\pm3.47$} & $\underline{82.26\pm9.98}$ & $7.79\pm3.98$ \\
\textbf{Ours}               & \checkmark & $\mathbf{95.17\pm2.04}$ & $\mathbf{4.20\pm1.70}$ & $\mathbf{83.25\pm16.67}$ & $\mathbf{6.16\pm3.34}$ \\
\hline
\end{tabular}
}
\label{tab:comparison_results}
\end{table*}

We then define a binary mask $m_v^{\text{quantile}}$ to select predictions that fall between these thresholds:
\begin{align}
    m_v^{\text{quantile}} = \mathds{1}[p_v^S > q_{\text{low}}] \cdot \mathds{1}[p_v^S < q_{\text{high}}].
\end{align}
This mask filters out extremely high-confidence foreground and background predictions. Entropy minimization is applied only to the selected uncertain regions:
\begin{align}
    \mathcal{L}_{\text{ent}} = - \sum_v m_v^{\text{quantile}} \cdot p^S_v \log p^S_v,
\end{align}
where $p^S_v$ denotes the sigmoid output of the student model at pixel $v$. By targeting only ambiguous areas that can benefit from additional supervision, this loss prevents the model from over-penalizing already confident regions.

The final objective function combines the consistency loss with the filtered entropy loss:
\begin{align}
    \mathcal{L}_\text{SFDA} = \mathcal{L}_{\text{cons}} + \mathcal{L}_{\text{ent}}.
\end{align}

%% file: sec/exp.tex
\section{Experiments}
\subsection{Implementation Details}
For fair comparison, we follow the same settings as \cite{DPL}, using a DeepLabV3+ model with a MobileNetV2 backbone. The output threshold $\gamma$ is 0.75. We use the Adam optimizer with a learning rate of $1 \times 10^{-3}$ for 200 epochs on the source domain. During source-free adaptation, we train for 20 epochs with a learning rate of $5 \times 10^{-4}$. Both teacher and student models are initialized from the pretrained source model, with Uncertainty-Guided EMA applied between them using $\alpha=0.95$. We set $\eta_1=0.05$, perform 10 stochastic forward passes, and define $\eta_2$ as the median of foreground and background entropy. The quantile threshold $\beta$ is 0.1, and the Gaussian scaling factor $s$ is 0.25. Strong data augmentations (contrast adjustment, random erasing, Gaussian noise) are applied. The implementation is based on PyTorch and runs on a single NVIDIA 3090Ti GPU.

\begin{figure*}[!ht]
    \centering
    \renewcommand{\arraystretch}{1.15}
    \setlength{\tabcolsep}{2pt}
    \resizebox{1\textwidth}{!}{%
    \begin{tabular}{rccccccccc}
        \multirow{2}{*}{\rotatebox[origin=c]{90}{\scriptsize Drishti-G}} &
        \includegraphics[width=0.09\linewidth]{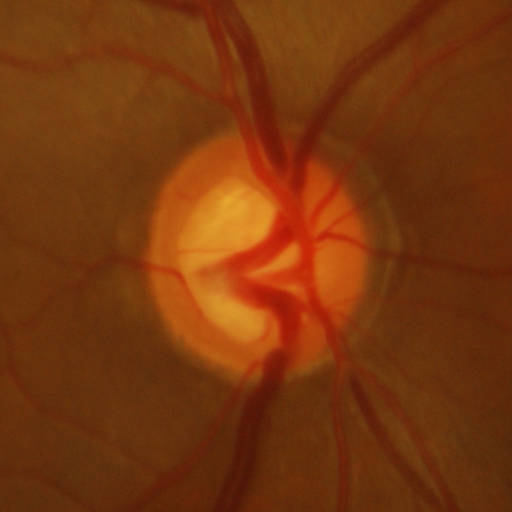} &
        \includegraphics[width=0.09\linewidth]{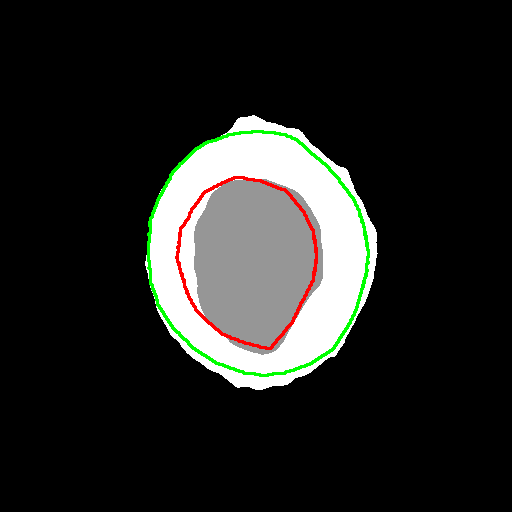} &
        \includegraphics[width=0.09\linewidth]{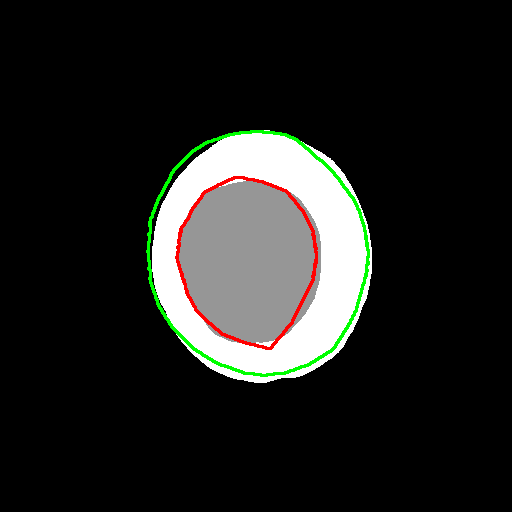} &
        \includegraphics[width=0.09\linewidth]{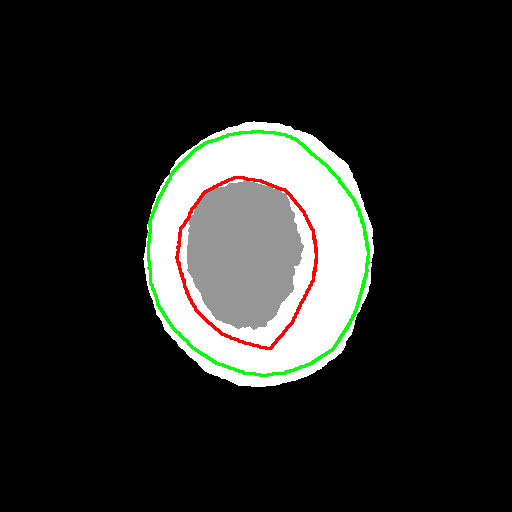} &
        \includegraphics[width=0.09\linewidth]{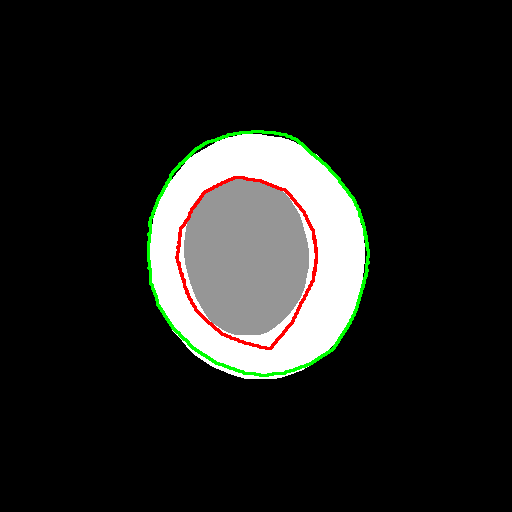} &
        \includegraphics[width=0.09\linewidth]{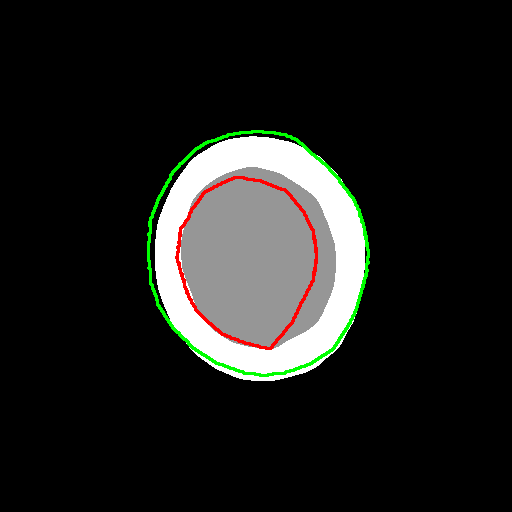} &
        \includegraphics[width=0.09\linewidth]{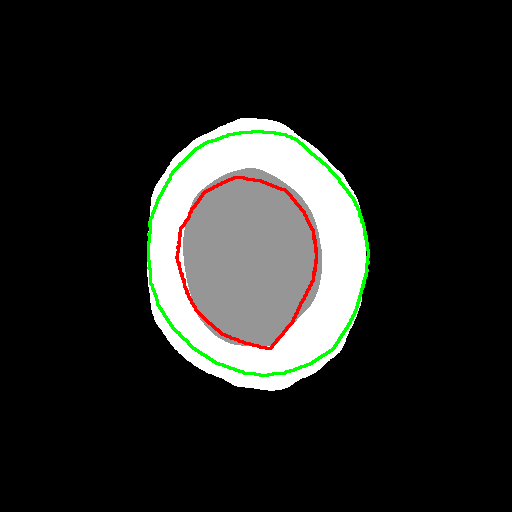} &
        \includegraphics[width=0.09\linewidth]{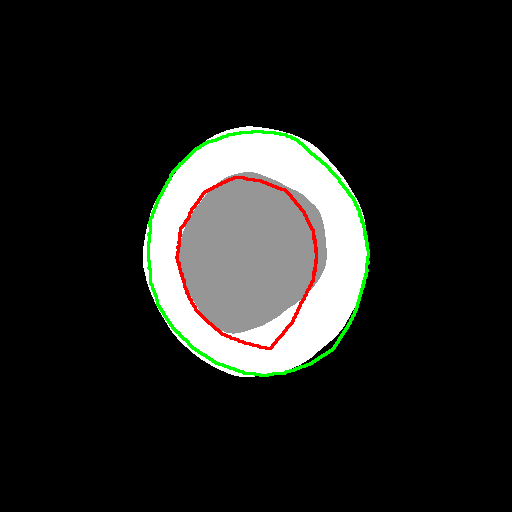} &
        \includegraphics[width=0.09\linewidth]{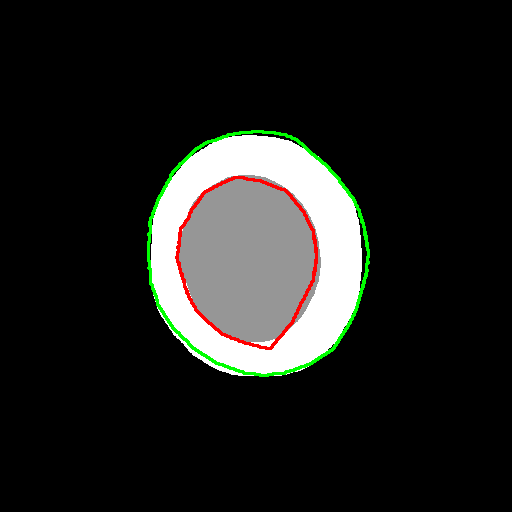} \\
        & \includegraphics[width=0.09\linewidth]{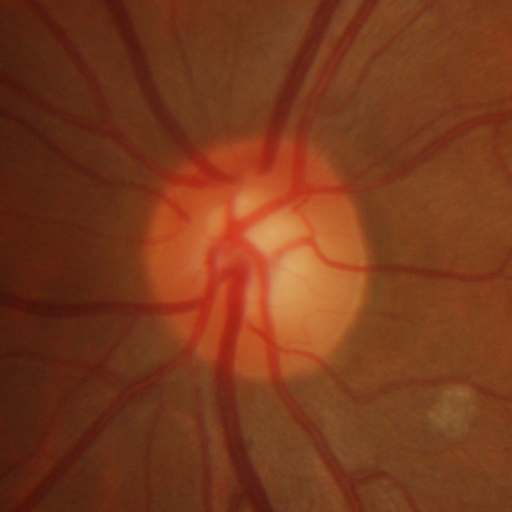} &
        \includegraphics[width=0.09\linewidth]{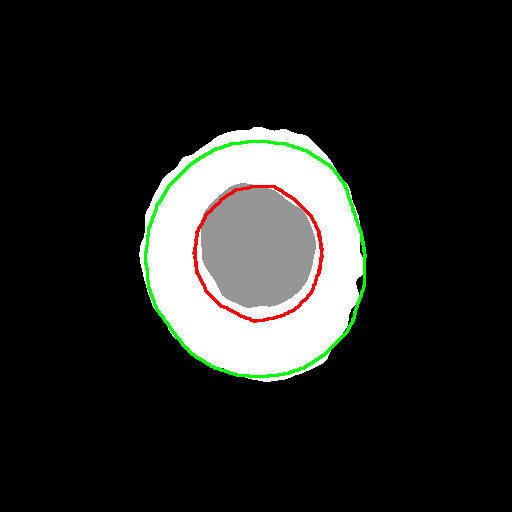} &
        \includegraphics[width=0.09\linewidth]{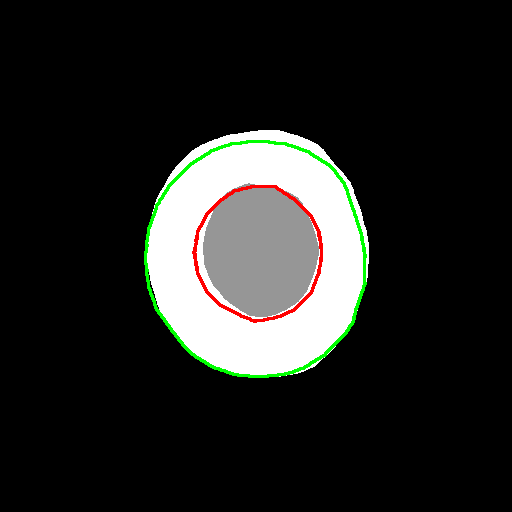} &
        \includegraphics[width=0.09\linewidth]{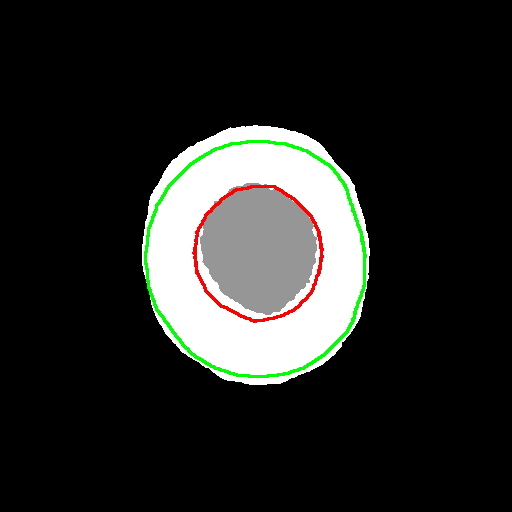} &
        \includegraphics[width=0.09\linewidth]{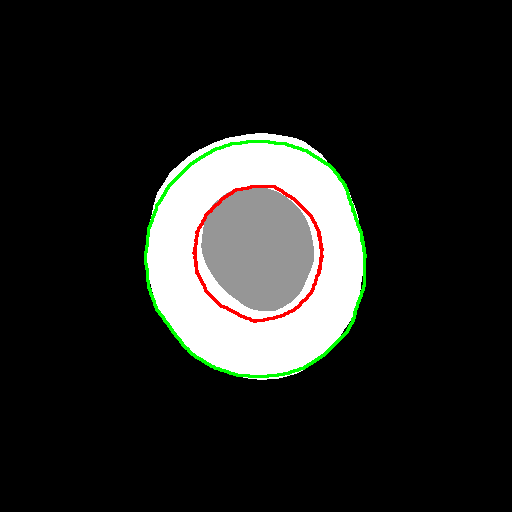} &
        \includegraphics[width=0.09\linewidth]{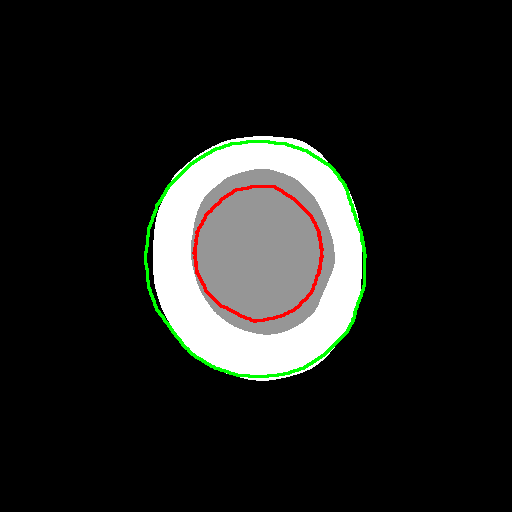} &
        \includegraphics[width=0.09\linewidth]{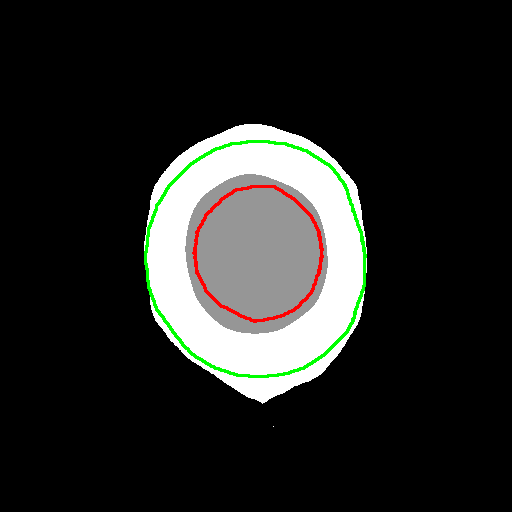} &
        \includegraphics[width=0.09\linewidth]{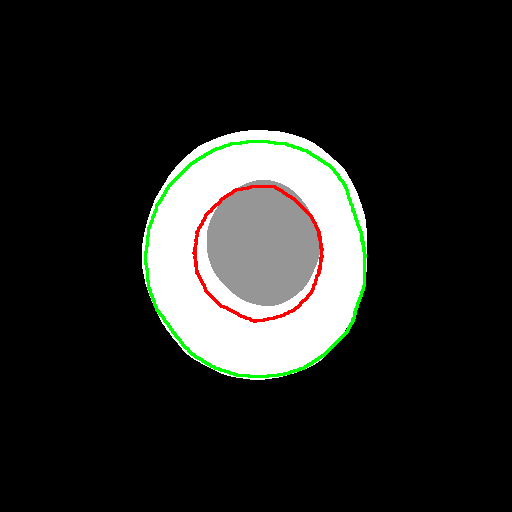} &
        \includegraphics[width=0.09\linewidth]{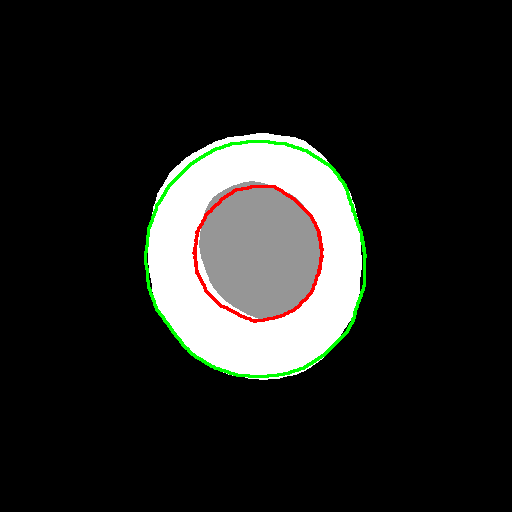} \\

        \multirow{2}{*}{\rotatebox[origin=c]{90}{\scriptsize RIM-ONE}} &
        \includegraphics[width=0.09\linewidth]{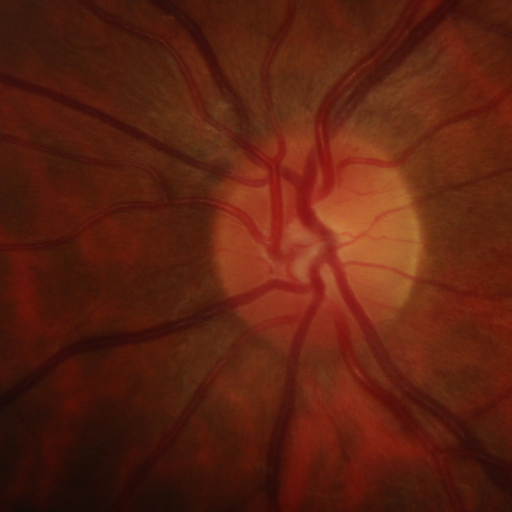} &
        \includegraphics[width=0.09\linewidth]{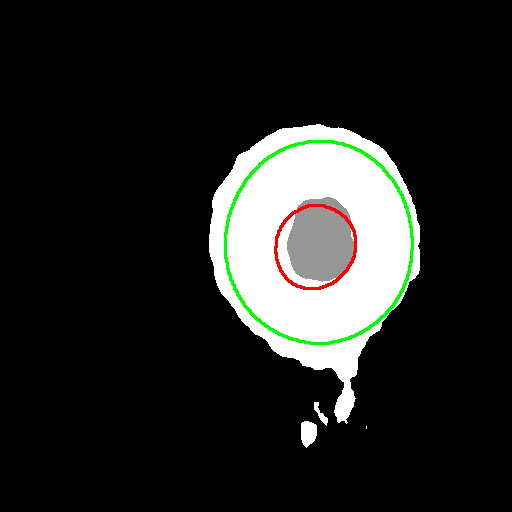} &
        \includegraphics[width=0.09\linewidth]{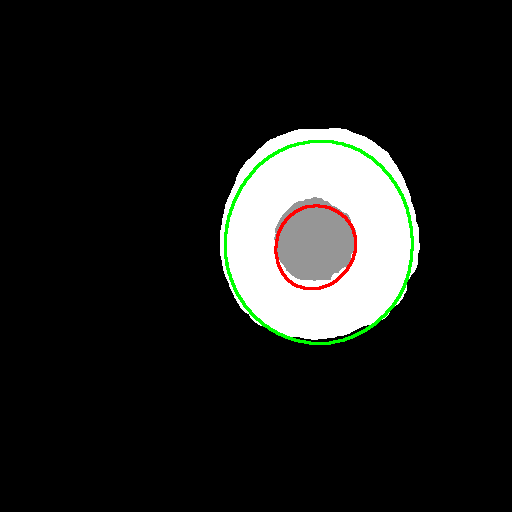} &
        \includegraphics[width=0.09\linewidth]{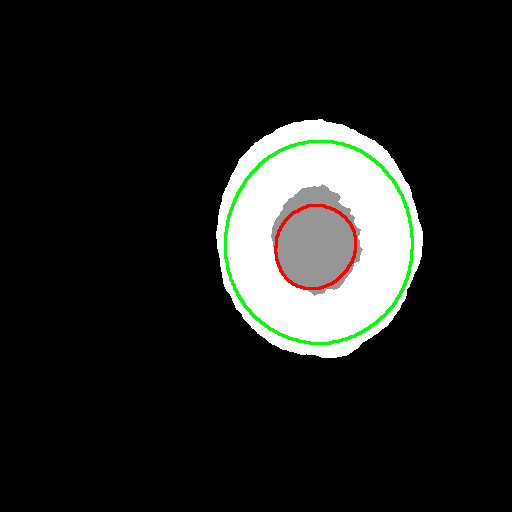} &
        \includegraphics[width=0.09\linewidth]{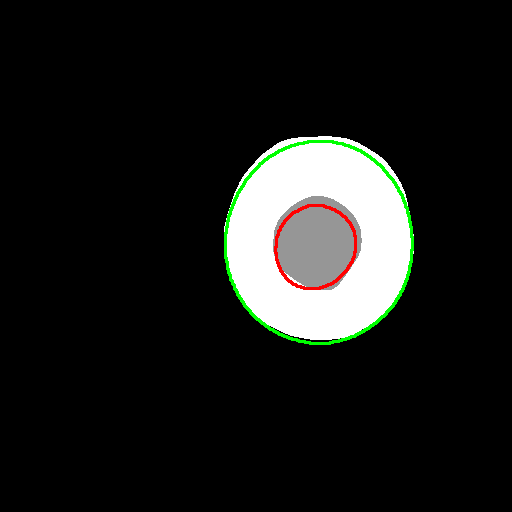} &
        \includegraphics[width=0.09\linewidth]{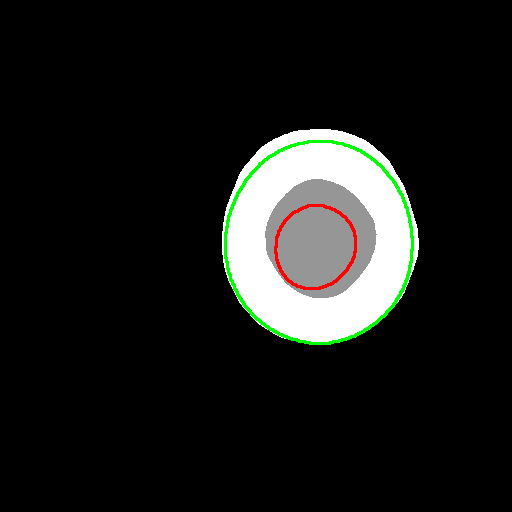} &
        \includegraphics[width=0.09\linewidth]{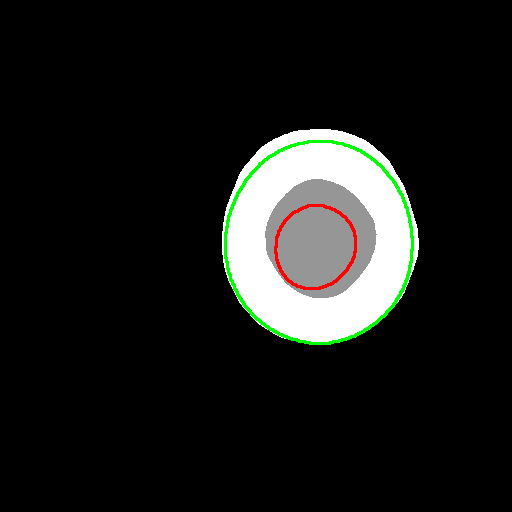} &
        \includegraphics[width=0.09\linewidth]{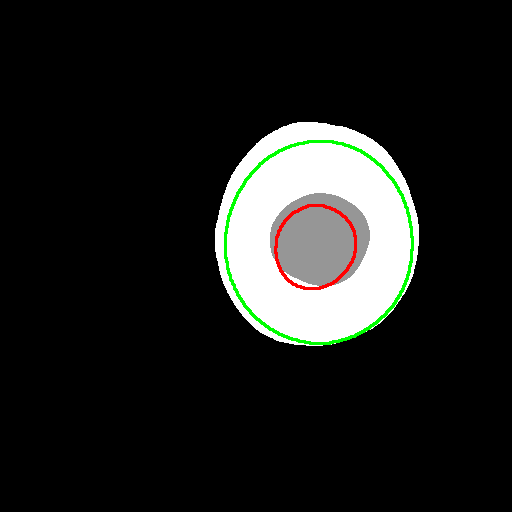} &
        \includegraphics[width=0.09\linewidth]{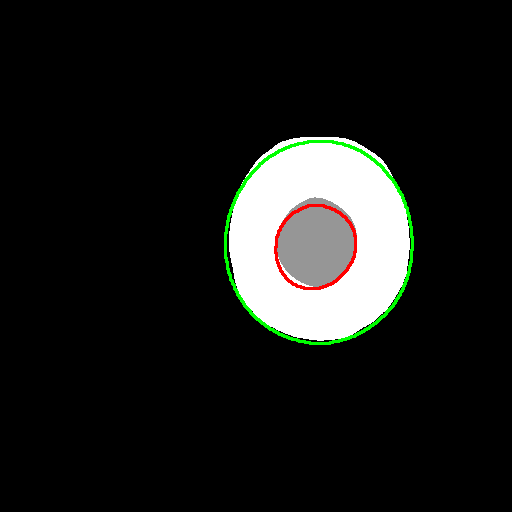} \\
        & \includegraphics[width=0.09\linewidth]{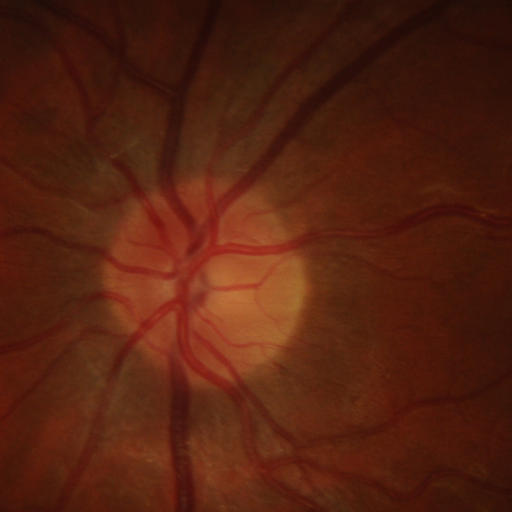} &
        \includegraphics[width=0.09\linewidth]{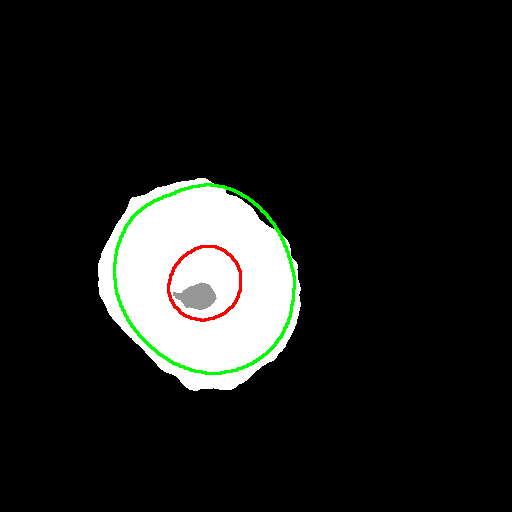} &
        \includegraphics[width=0.09\linewidth]{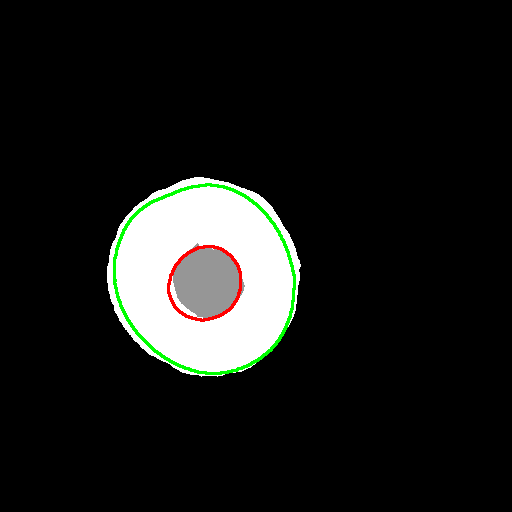} &
        \includegraphics[width=0.09\linewidth]{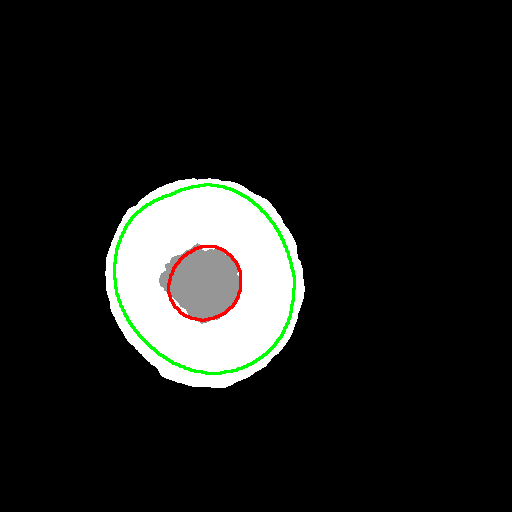} &
        \includegraphics[width=0.09\linewidth]{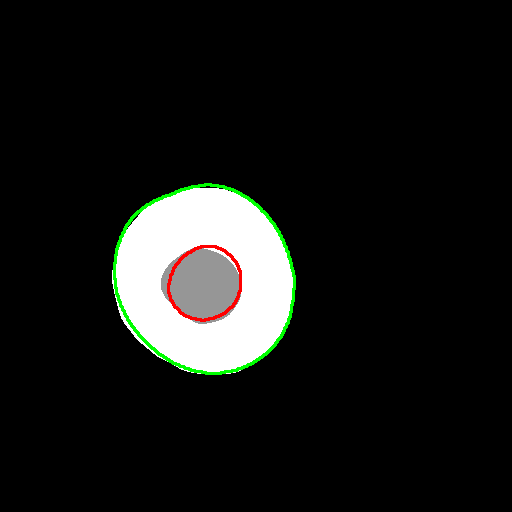} &
        \includegraphics[width=0.09\linewidth]{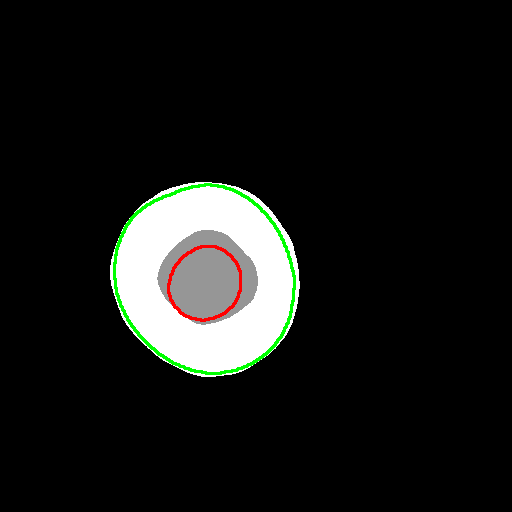} &
        \includegraphics[width=0.09\linewidth]{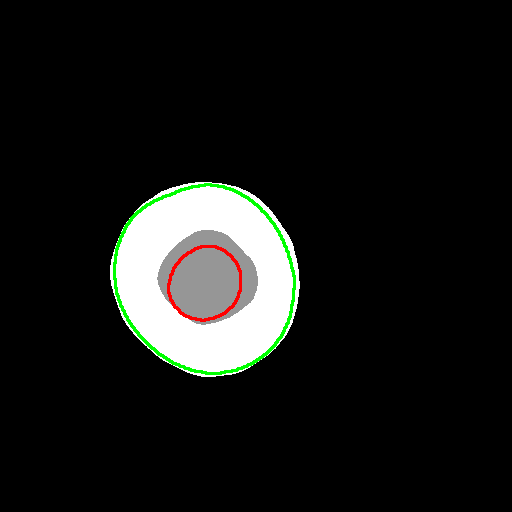} &
        \includegraphics[width=0.09\linewidth]{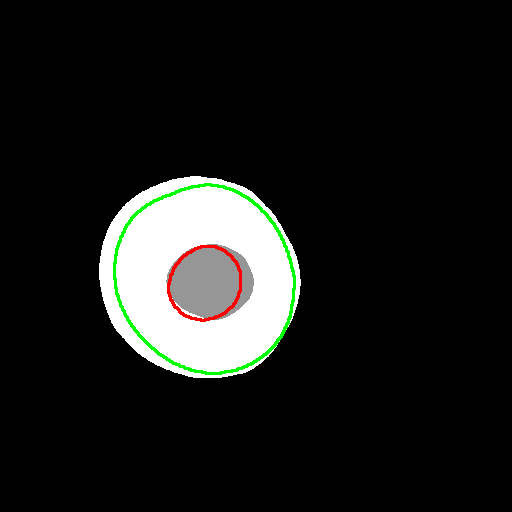} &
        \includegraphics[width=0.09\linewidth]{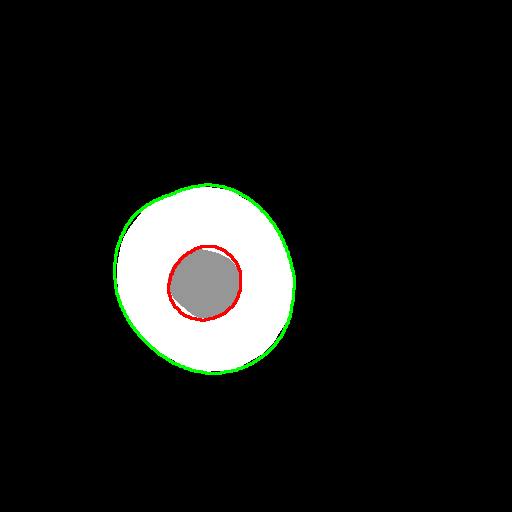} \\

        \multirow{2}{*}{\rotatebox[origin=c]{90}{\scriptsize Open}} &
        \includegraphics[width=0.09\linewidth]{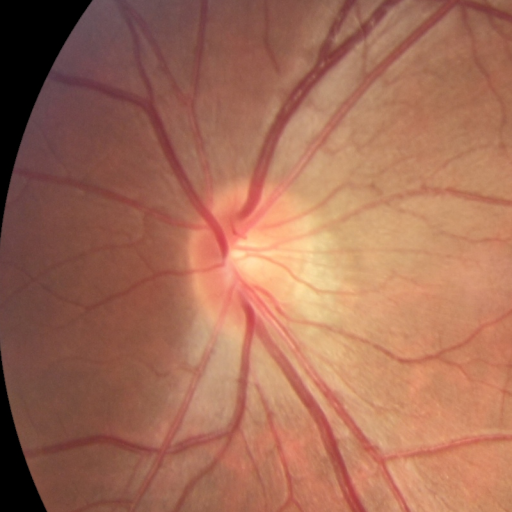} &
        \includegraphics[width=0.09\linewidth]{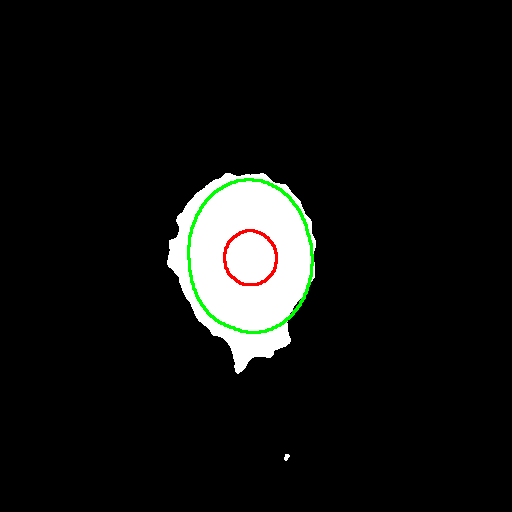} &
        \includegraphics[width=0.09\linewidth]{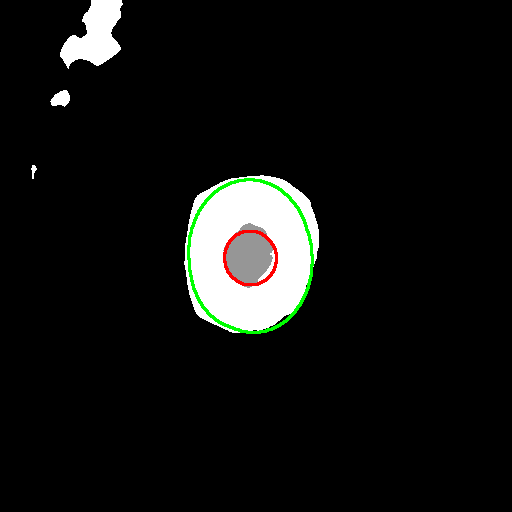} &
        \includegraphics[width=0.09\linewidth]{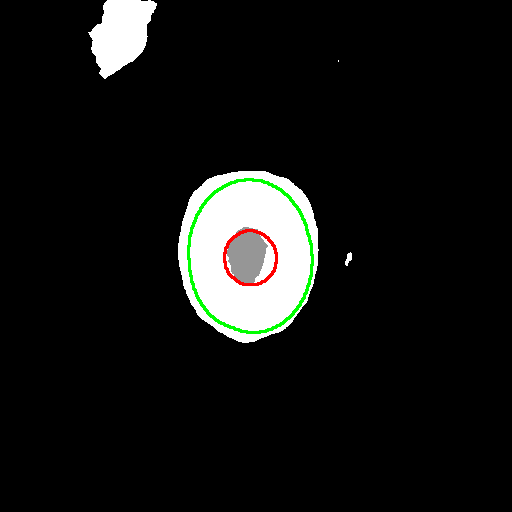} &
        \includegraphics[width=0.09\linewidth]{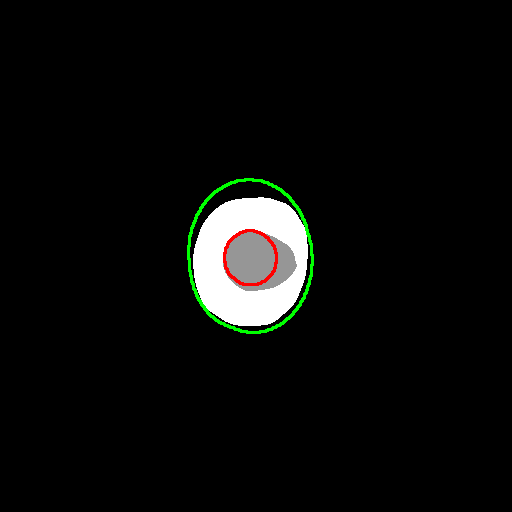} &
        \includegraphics[width=0.09\linewidth]{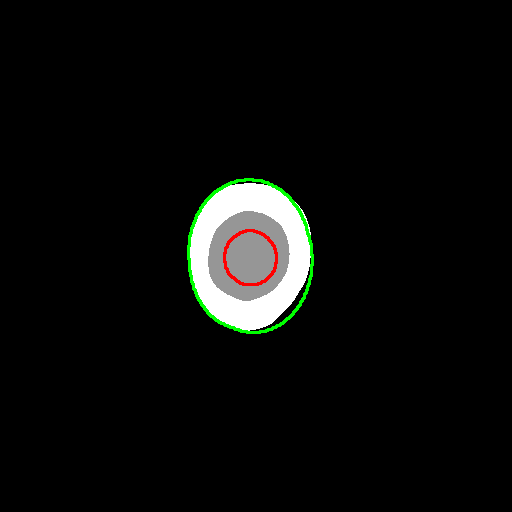} &
        \includegraphics[width=0.09\linewidth]{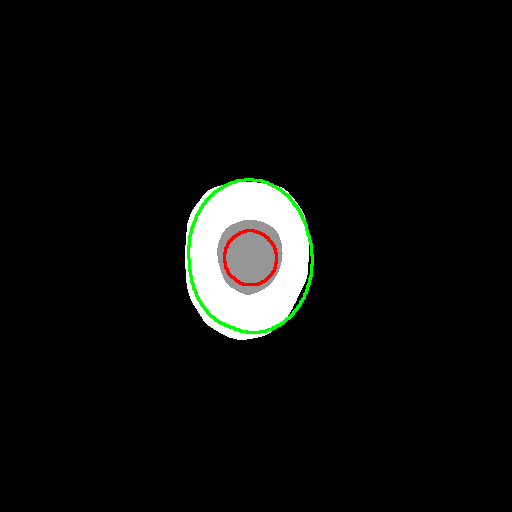} &
        \includegraphics[width=0.09\linewidth]{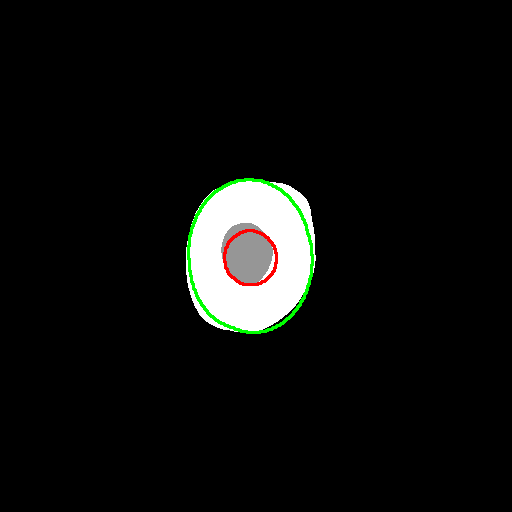} &
        \includegraphics[width=0.09\linewidth]{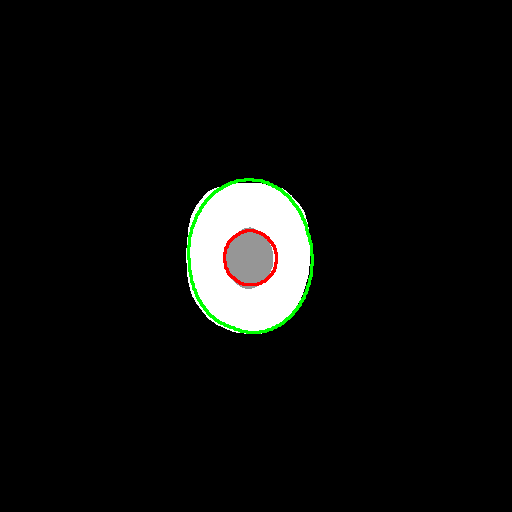} \\
        & \includegraphics[width=0.09\linewidth]{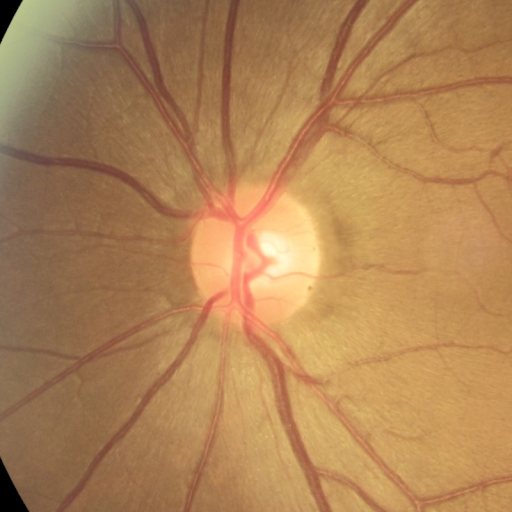} &
        \includegraphics[width=0.09\linewidth]{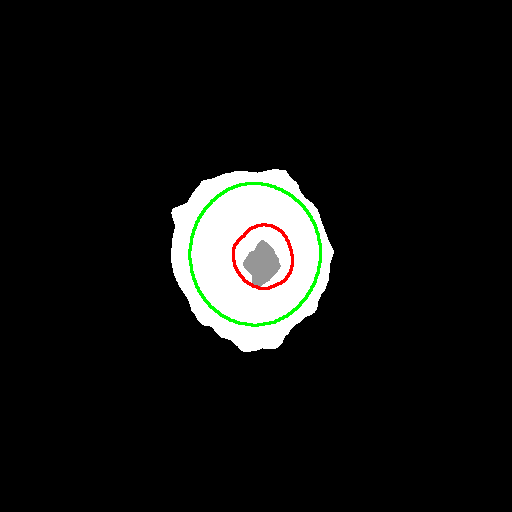} &
        \includegraphics[width=0.09\linewidth]{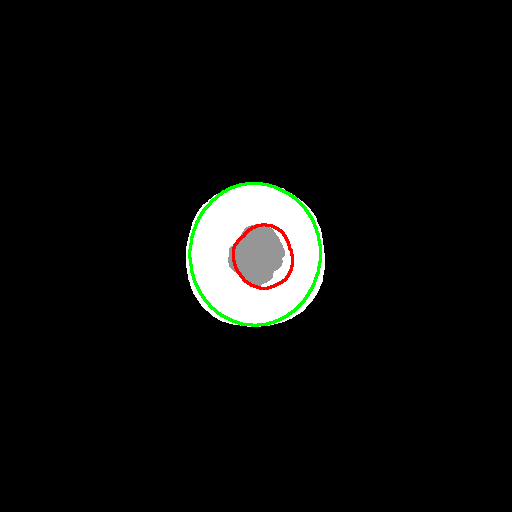} &
        \includegraphics[width=0.09\linewidth]{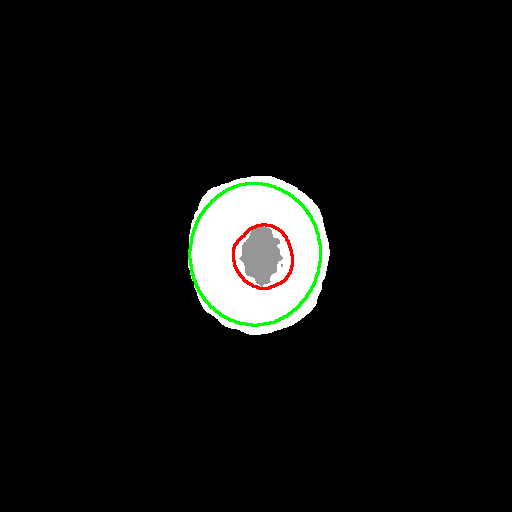} &
        \includegraphics[width=0.09\linewidth]{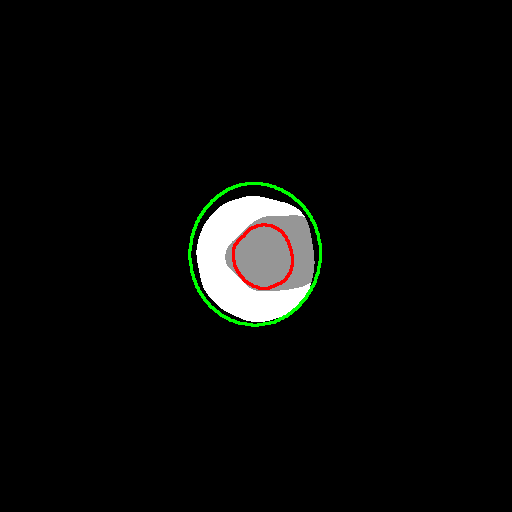} &
        \includegraphics[width=0.09\linewidth]{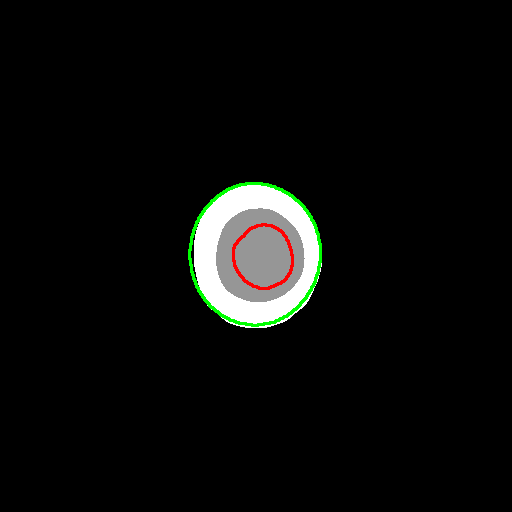} &
        \includegraphics[width=0.09\linewidth]{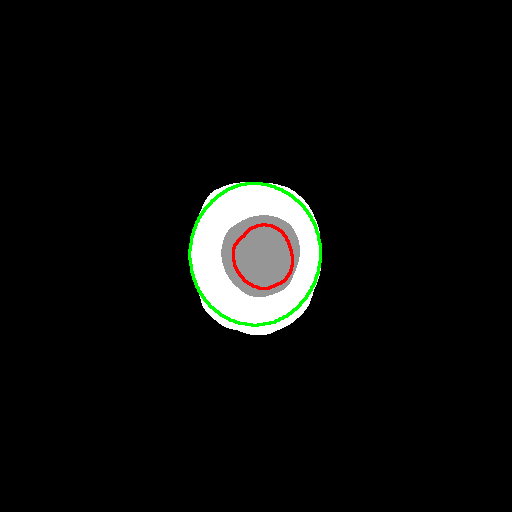} &
        \includegraphics[width=0.09\linewidth]{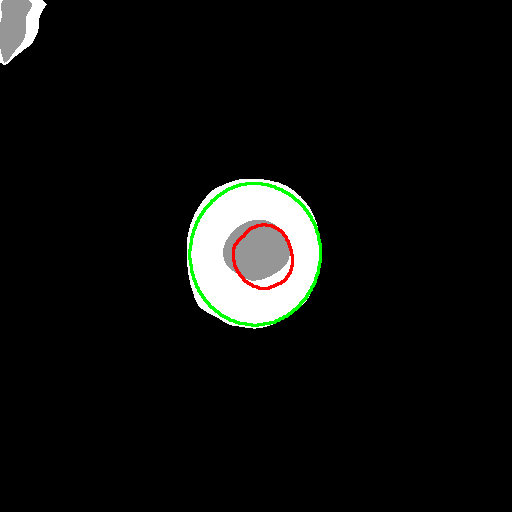} &
        \includegraphics[width=0.09\linewidth]{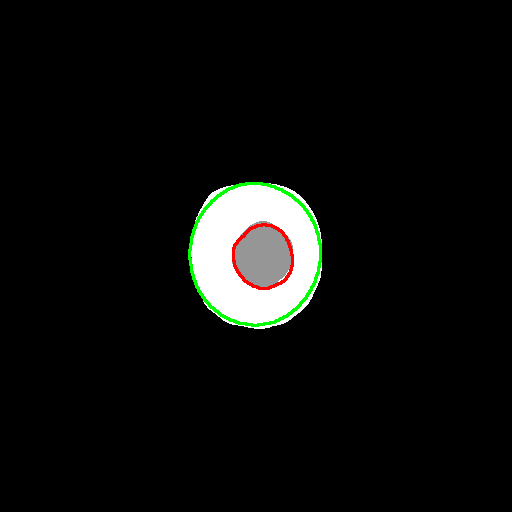} \\

        & \footnotesize Image 
        & \footnotesize AdvEnt 
        & \footnotesize BEAL 
        & \footnotesize TENT 
        & \footnotesize CBMT 
        & \footnotesize CPR 
        & \footnotesize DPL 
        & \footnotesize PLPB 
        & \footnotesize \textbf{Ours} \\
    \end{tabular}
    }
    \caption{Qualitative comparisons of different methods with REFUGE as source domain.}
    \label{fig:qualitative_results}
\end{figure*}

\begin{table*}[!ht]
\centering
\caption{Quantitative comparison of results on DrishtiGS under SFDA settings using different metrics for UG-EMA method. The best score in each column is indicated in bold, while the second-best score is underlined.}
\renewcommand{\arraystretch}{1.2}
\resizebox{\textwidth}{!}{%
\begin{tabular}{|c|c|c|c|c|c|c|}
\hline
\multirow{2}{*}{\textbf{Metrics}} &
\multicolumn{2}{c|}{\textbf{Optic Disc Segmentation}} &
\multicolumn{2}{c|}{\textbf{Optic Cup Segmentation}} &
\multicolumn{2}{c|}{\textbf{Avg}} \\
\cline{2-7}
& \textbf{Dice[\%] $\uparrow$} & \textbf{ASSD[pixel] $\downarrow$} 
& \textbf{Dice[\%] $\uparrow$} & \textbf{ASSD[pixel] $\downarrow$}
& \textbf{Dice[\%] $\uparrow$} & \textbf{ASSD[pixel] $\downarrow$} \\
\hline
Loss & 96.51 $\pm$ 1.26 & 3.89 $\pm$ 1.34 & \underline{86.08 $\pm$ 11.92} & \underline{9.12 $\pm$ 5.41} & \underline{91.29} & \underline{6.51} \\
\hline
Full entropy & \underline{96.55 $\pm$ 1.35} & \underline{3.88 $\pm$ 1.50} & 85.66 $\pm$ 13.13 & 9.69 $\pm$ 6.29 & 91.11 & 6.78 \\
\hline
\textbf{Entropy with Inverted Gaussian Weight} & 
\textbf{96.58 $\pm$ 1.27} & \textbf{3.86 $\pm$ 1.40} & \textbf{86.71 $\pm$ 12.43} & \textbf{8.69 $\pm$ 5.41}  & \textbf{91.64} & \textbf{6.28} \\
\hline

\end{tabular}%
}
\label{tab:metrics}
\end{table*}

\subsection{Datasets and Metrics}
We evaluate on widely used datasets for optic disc and cup segmentation. Following previous studies, REFUGE \cite{orlando2020refuge} is used as the source domain, while RIM-ONE-r3 \cite{fumero2011rim}, Drishti-GS \cite{sivaswamy2015comprehensive}, and REFUGE validation \cite{orlando2020refuge} (open domain) are used as targets. The splits are 320/80 (REFUGE), 90/60 (RIM-ONE-r3), and 50/51 (Drishti-GS) for training/testing, with 80 open-domain images. As in \cite{BEAL}, fundus images are ROI-cropped to $512 \times 512$. We report Dice coefficient (DICE) and Average Symmetric Surface Distance (ASSD) as evaluation metrics.

\begin{table*}[!ht]
\centering
\caption{The quantitative results on Open compound settings using different methods. \textbf{Bold} texts highlight the best scores.}
\resizebox{1\textwidth}{!}{
\begin{tabular}{|l|c|c|c|c|c|c|c|c|c|c|}
\hline
\multirow{3}{*}{\textbf{Method}} & 
\multicolumn{4}{c|}{\textbf{Compound (C)}} & 
\multicolumn{2}{c|}{\textbf{Open (O)}} & 
\multicolumn{4}{c|}{\textbf{Avg.}} \\
\cline{2-11}
& \multicolumn{2}{c|}{Drishti-GS} & \multicolumn{2}{c|}{RIM-ONE-r3} & \multicolumn{2}{c|}{REFUGE val} & \multicolumn{2}{c|}{C} & \multicolumn{2}{c|}{C+O} \\
\cline{2-11}
& Dice & ASSD & Dice & ASSD & Dice & ASSD & Dice & ASSD & Dice & ASSD \\
\hline
BEAL (MICCAI'19) \cite{BEAL} & \underline{90.65} & \underline{7.07} & 83.17 & 8.07  & \underline{88.10} & 8.38 & 86.91 & 7.57 & \underline{87.51} & 7.98 \\
AdvEnt (CVPR'19) \cite{AdvEnt} & 86.74 & 10.78 & 69.55 & 17.44  & 77.55 & 9.97 & 78.15 & 14.11 & 77.85 & 12.04 \\
TENT (ICLR'21) \cite{Tent} & 86.29 & 11.16 & 80.17 & 16.41  & 74.20 & 33.86 & 83.23 & 13.79 & 78.72 & 23.83 \\
DPL (MICCAI'21) \cite{DPL} & 88.03 & 9.81 & 75.24 & 16.74  & 86.01 & \underline{6.61} & 81.64 & 13.28 & 84.13 & 9.78 \\
CPR (MICCAI'23) \cite{CPR} & 90.08 & 7.52 & 83.72 & 8.65  & 83.83 & 7.47 & 86.90 & 8.09 & 85.37 & \underline{7.78} \\
CBMT (MICCAI'23) \cite{CBMT} & 90.47 & 7.08 & 87.26 & 7.29  & 79.12 & 12.43 & 88.87 & 7.19 & 83.99 & 9.81 \\
PLPB (WACV'24) \cite{PLPB} & 89.09 & 8.90 & 78.57 & 15.30  & 84.93 & 21.13 & 83.83 & 12.10 & 84.38 & 16.62 \\
SBIF (ISBI'25) \cite{yaacovi2025source} & 90.53 & \underline{7.07} & \underline{88.04} & \underline{6.69}  & 85.15 &  10.76 & \underline{89.29} & \underline{6.88} & 87.22 & 8.82 \\
\textbf{Ours} & \textbf{91.64} & \textbf{6.28} & \textbf{89.21} & \textbf{5.18}  & \textbf{88.63} & \textbf{4.87} & \textbf{90.43} & \textbf{5.73} & \textbf{89.53} & \textbf{5.3} \\
\hline
\end{tabular}%
}

\label{tab:open_avg}
\end{table*}

\begin{figure*}[!ht]
    \centering
    \renewcommand{\arraystretch}{0.4}
    \setlength{\tabcolsep}{1pt}
    \resizebox{1\textwidth}{!}{%
    \begin{tabular}{r P{0.14\linewidth} P{0.14\linewidth} P{0.14\linewidth} P{0.14\linewidth} P{0.14\linewidth} P{0.14\linewidth}}

        \raisebox{1.\height}{\rotatebox[origin=c]{90}{\tiny gdrishtiGS\_002}} &
        \includegraphics[width=\linewidth]{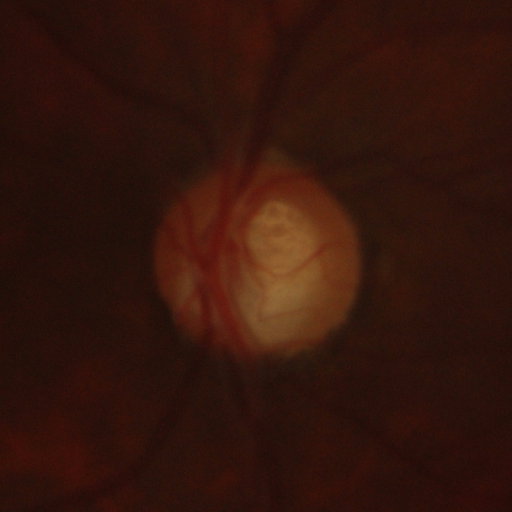} & 
        \includegraphics[width=\linewidth]{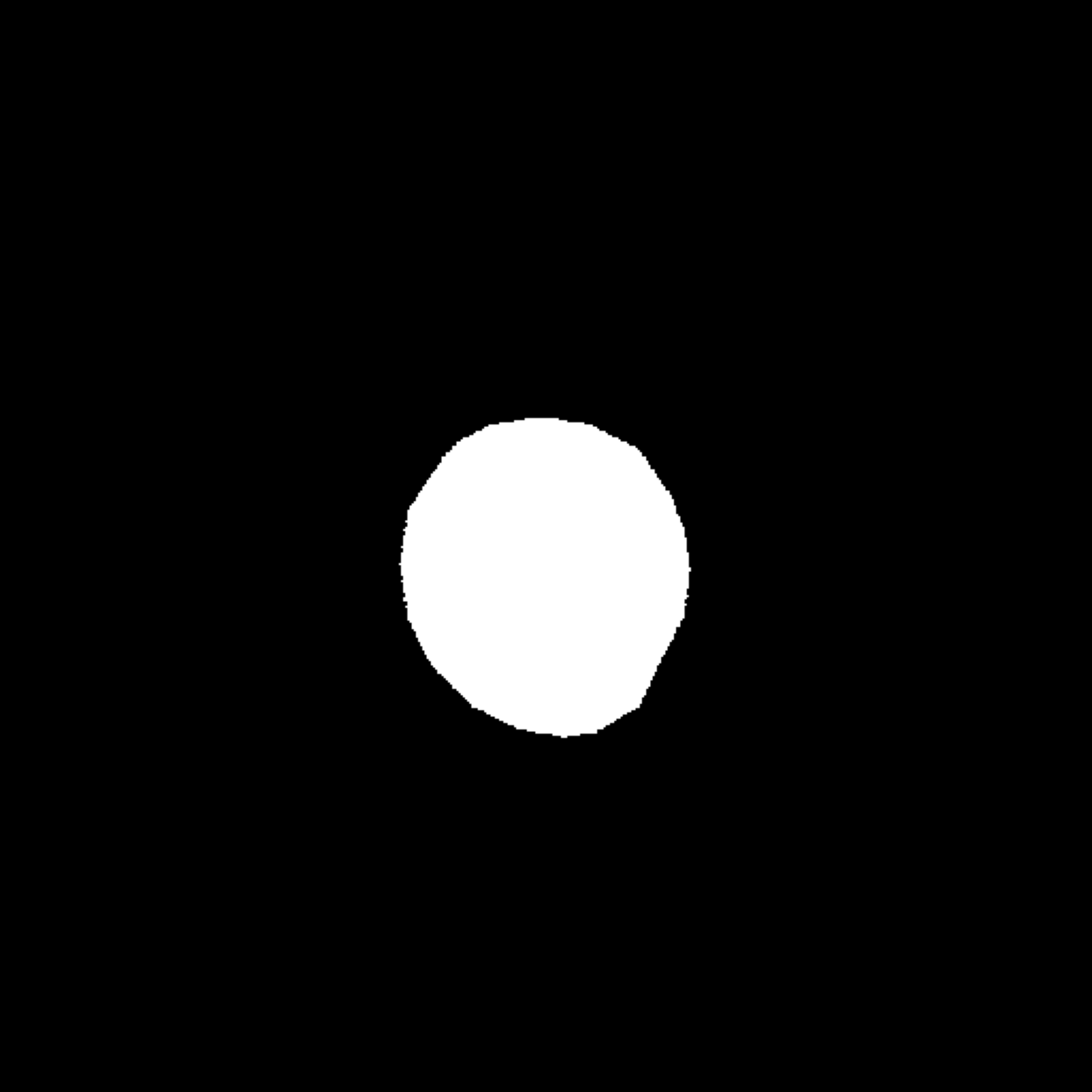} & 
        \includegraphics[width=\linewidth]{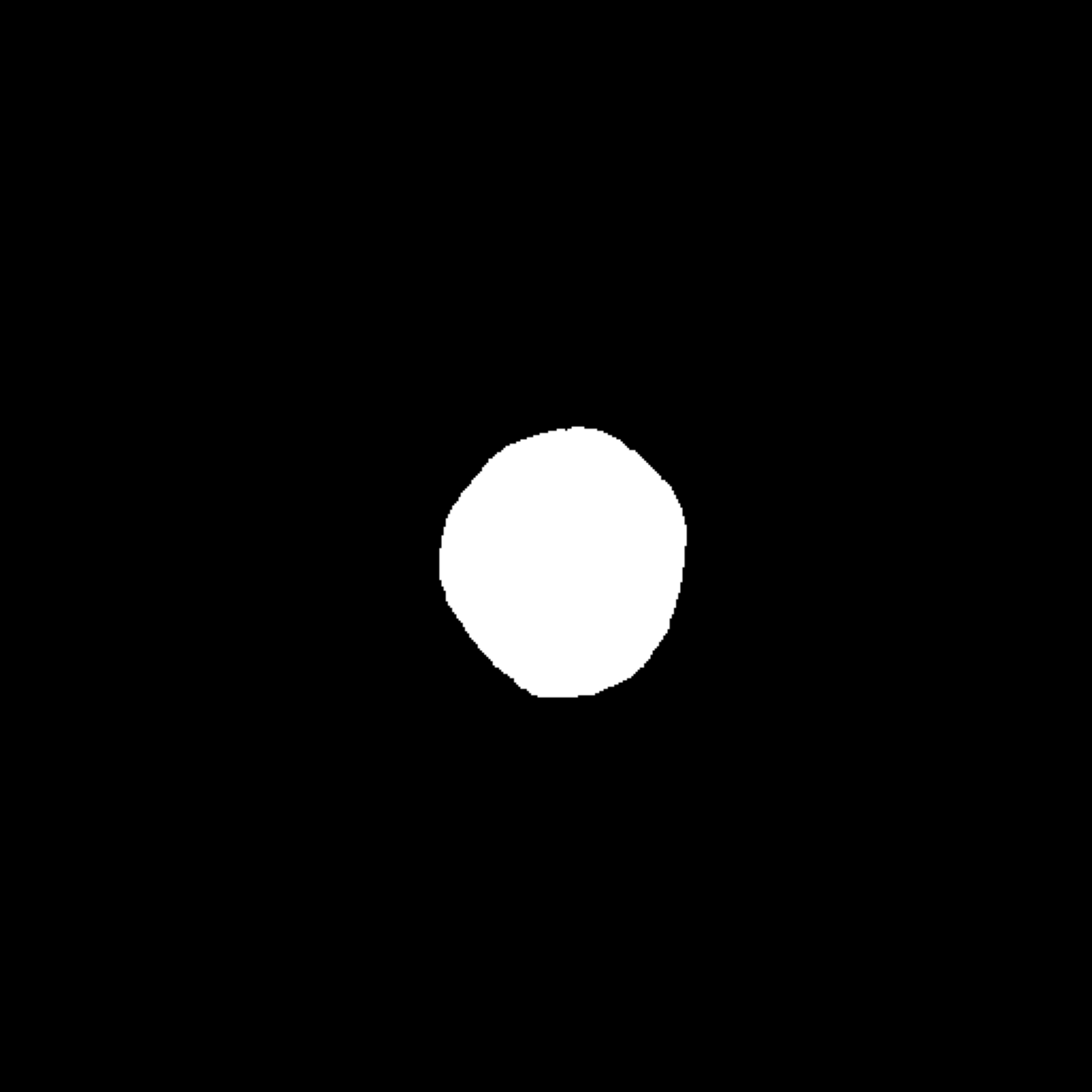} & 
        \includegraphics[width=\linewidth]{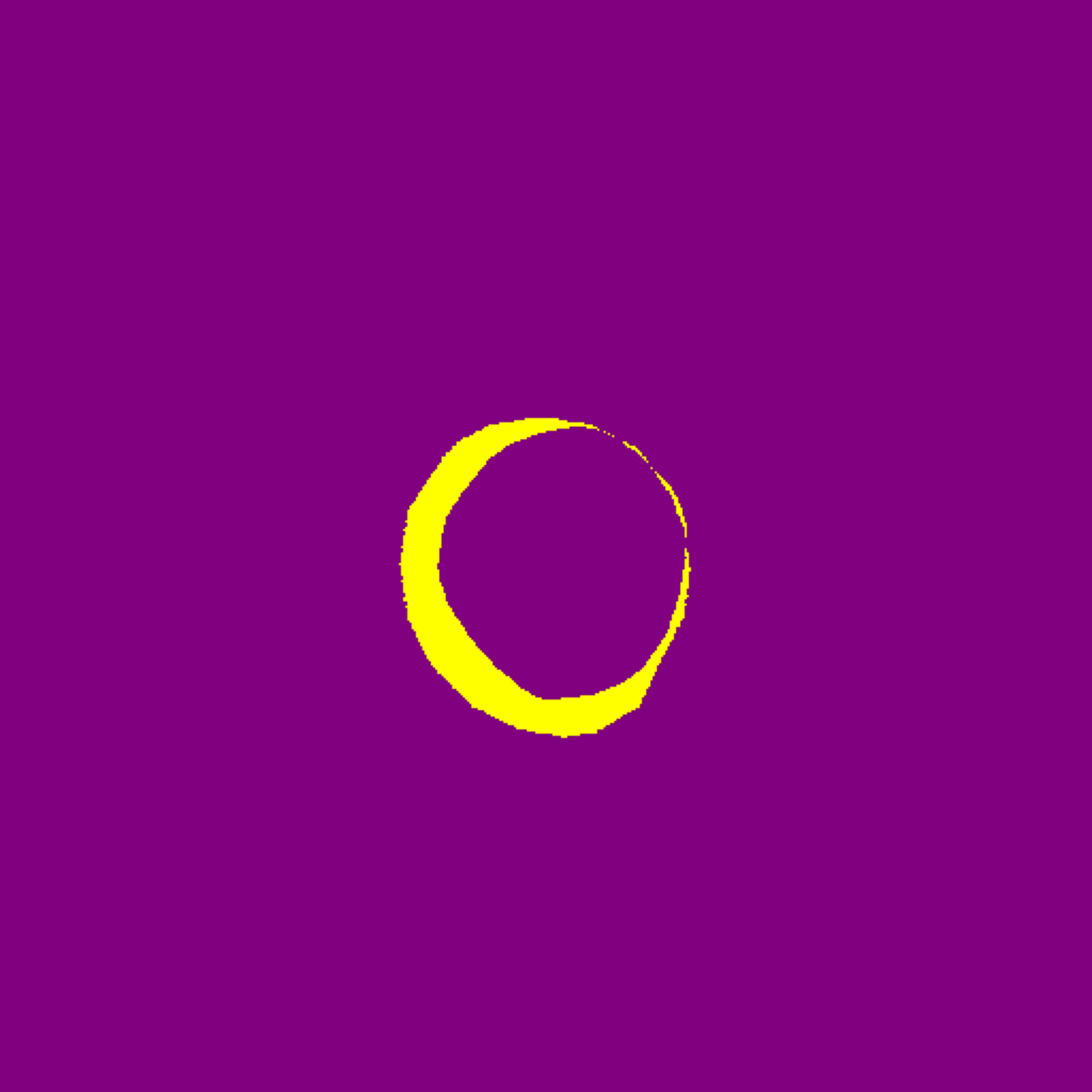} &
        \includegraphics[width=\linewidth]{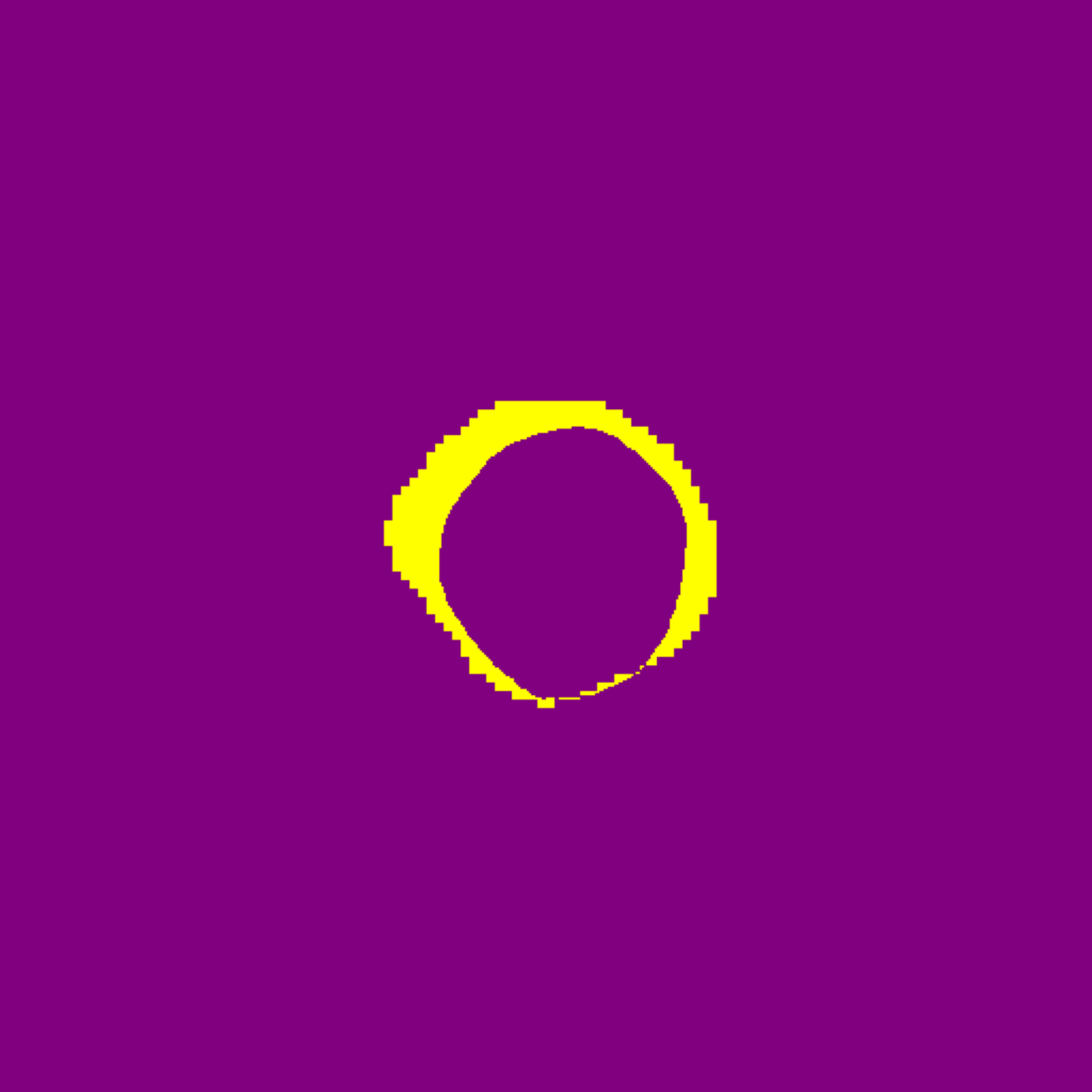} & 
        \includegraphics[width=\linewidth]{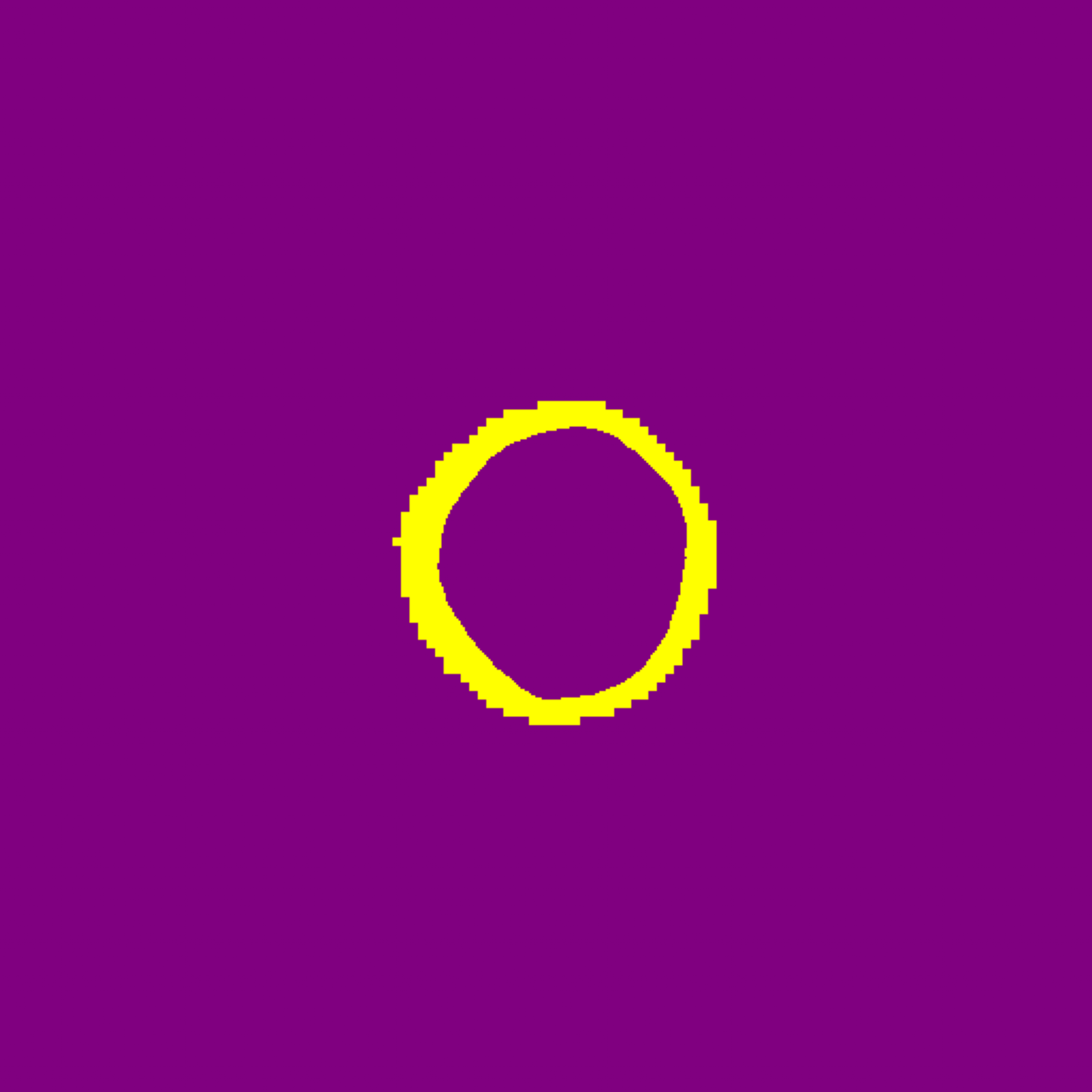} \\
        
        \raisebox{0.9\height}{\rotatebox[origin=c]{90}{\tiny ngdrishtiGS\_037}} &
        \includegraphics[width=\linewidth]{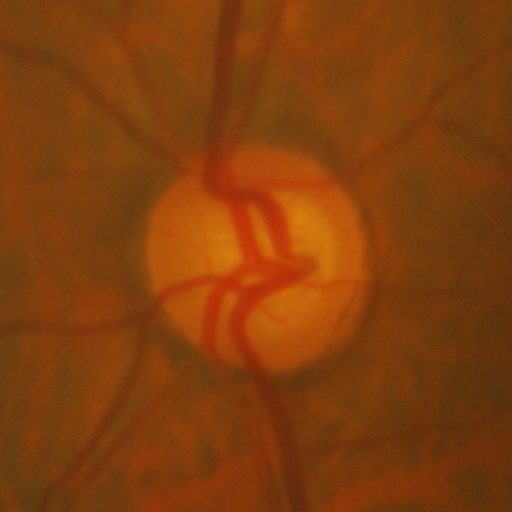} & 
        \includegraphics[width=\linewidth]{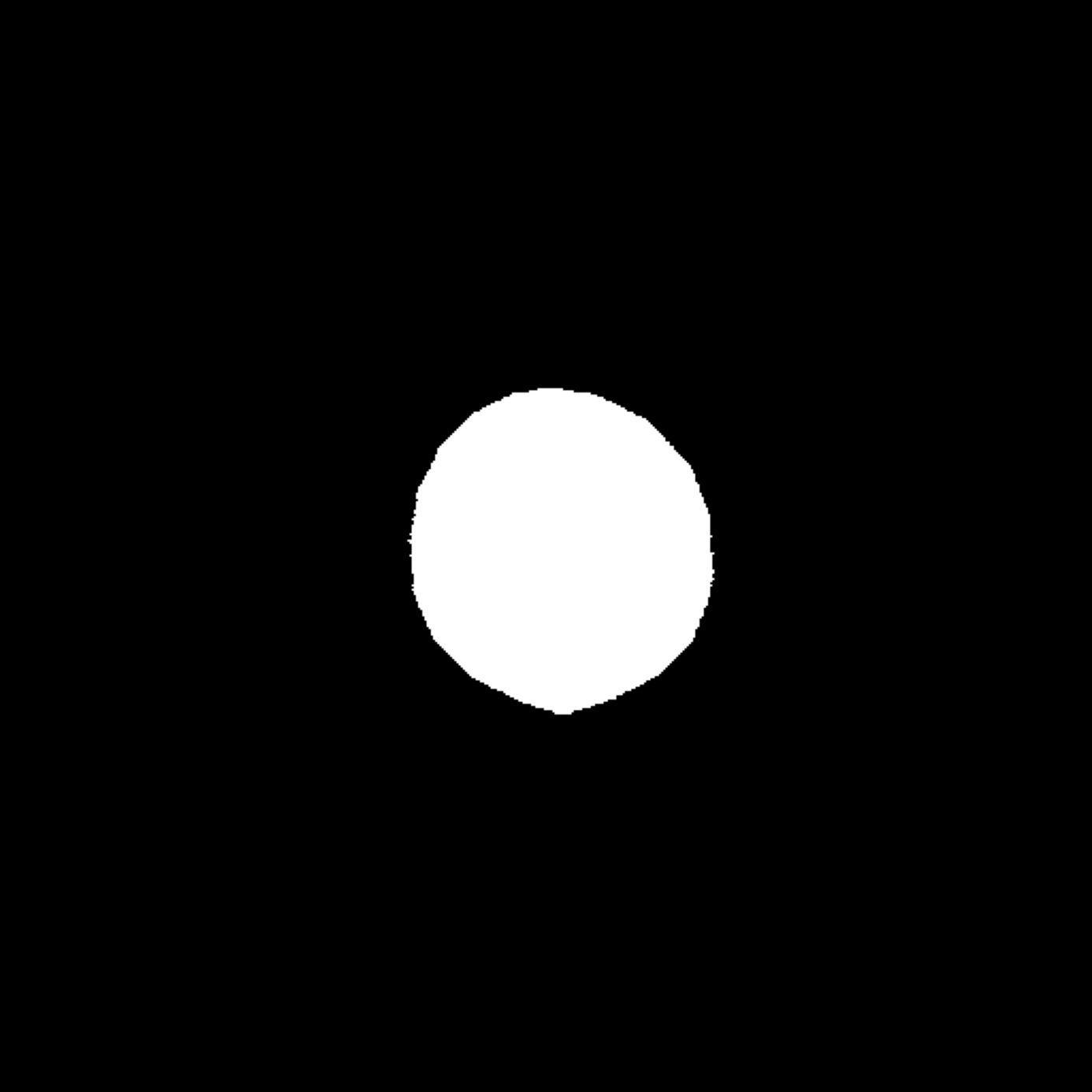} & 
        \includegraphics[width=\linewidth]{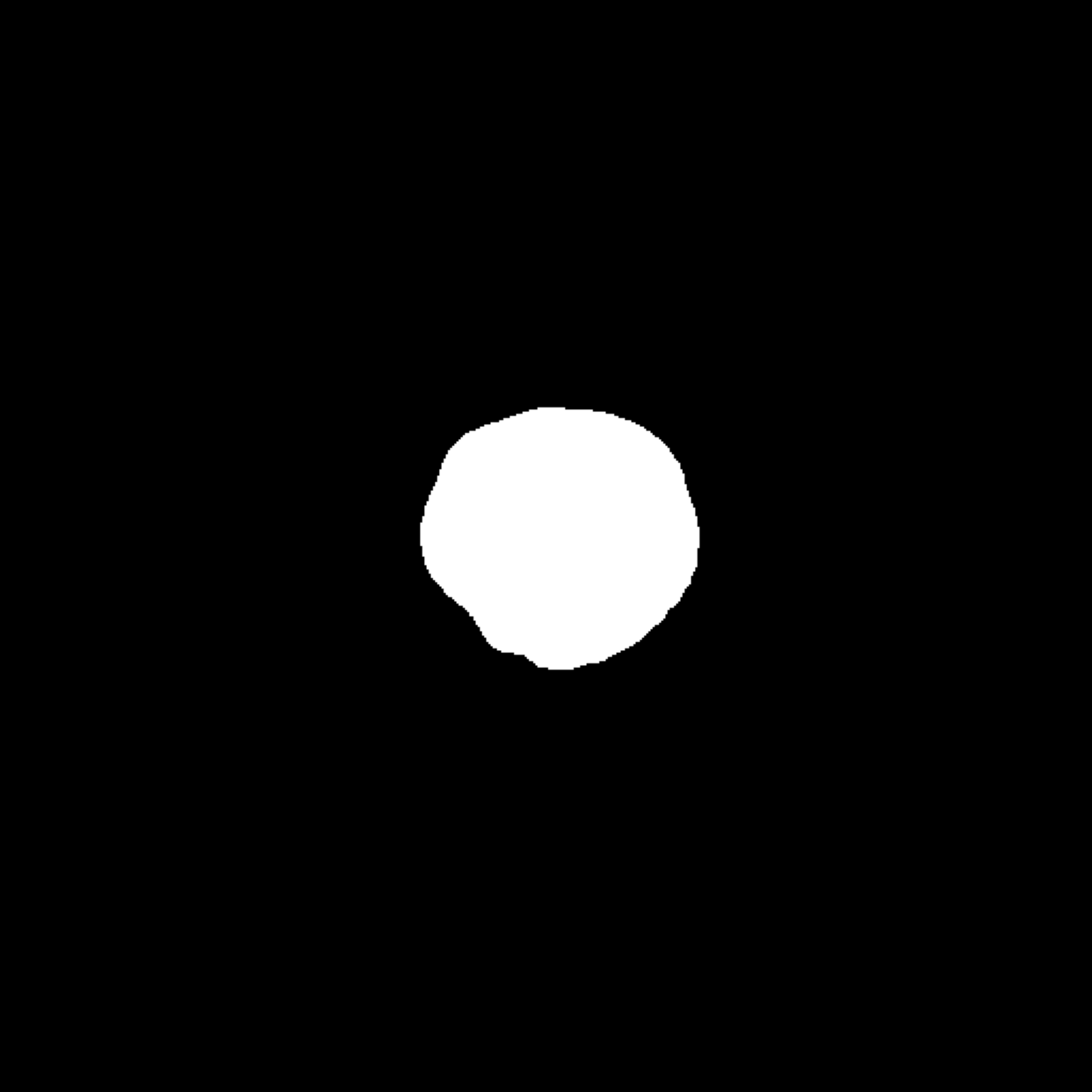} & 
        \includegraphics[width=\linewidth]{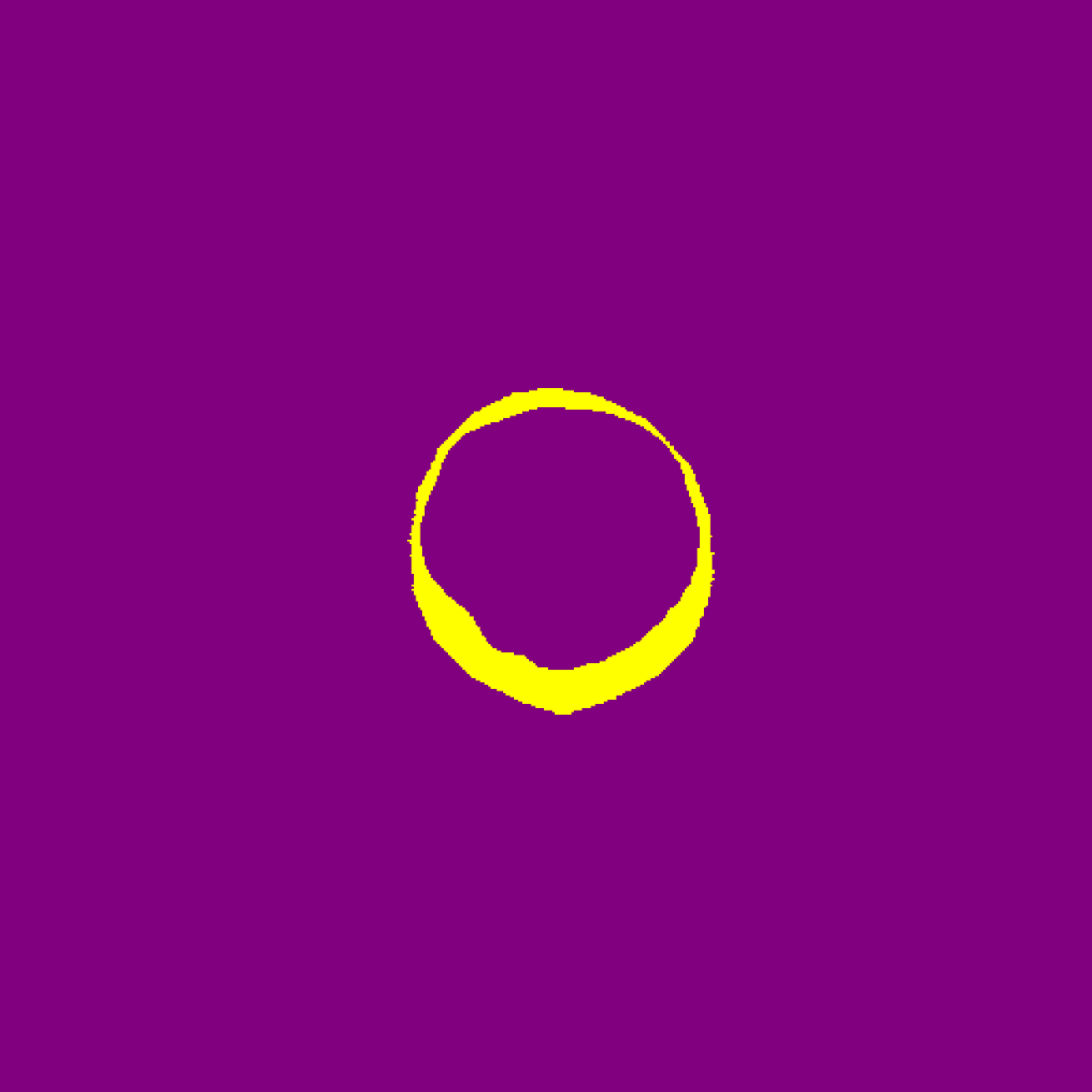} &
        \includegraphics[width=\linewidth]{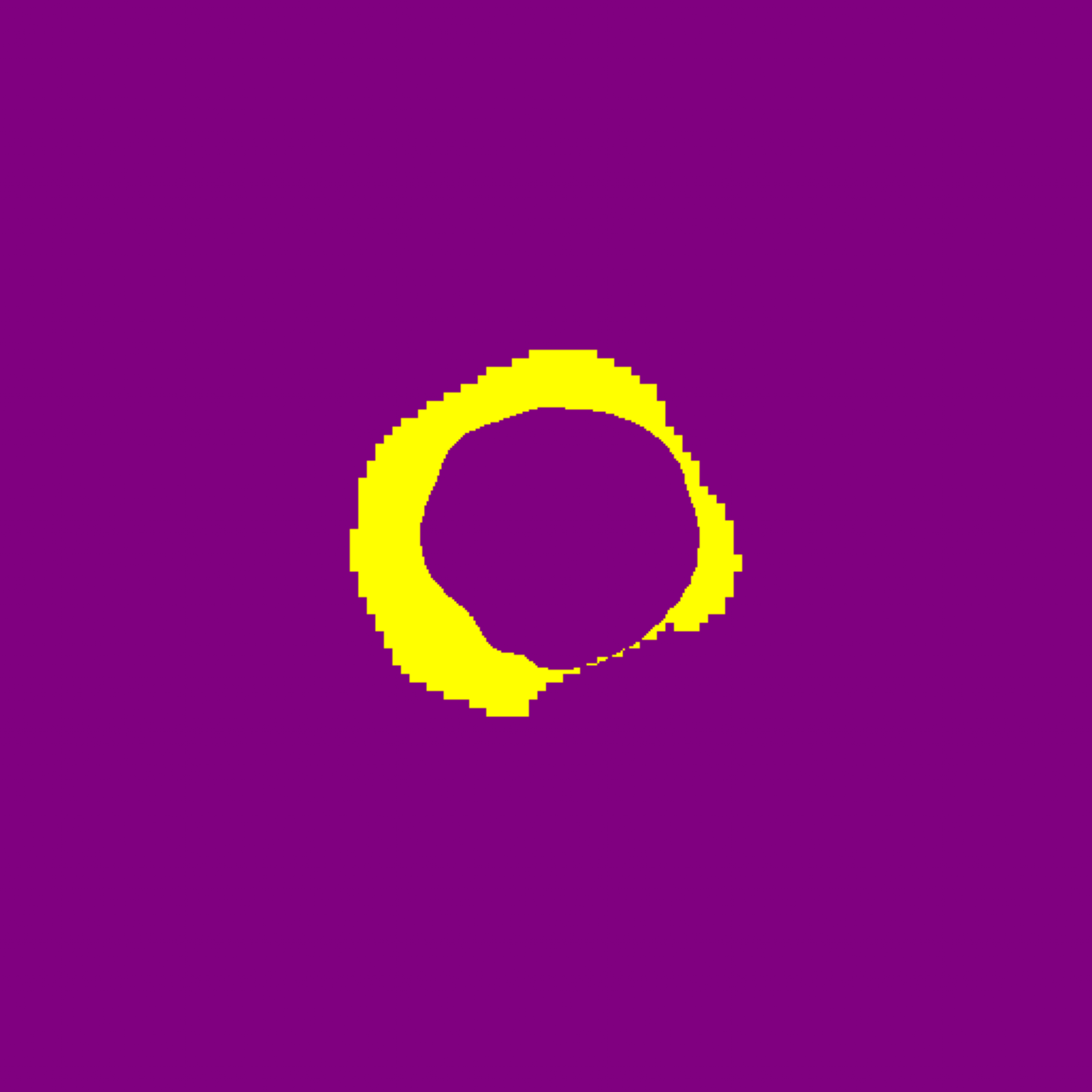} & 
        \includegraphics[width=\linewidth]{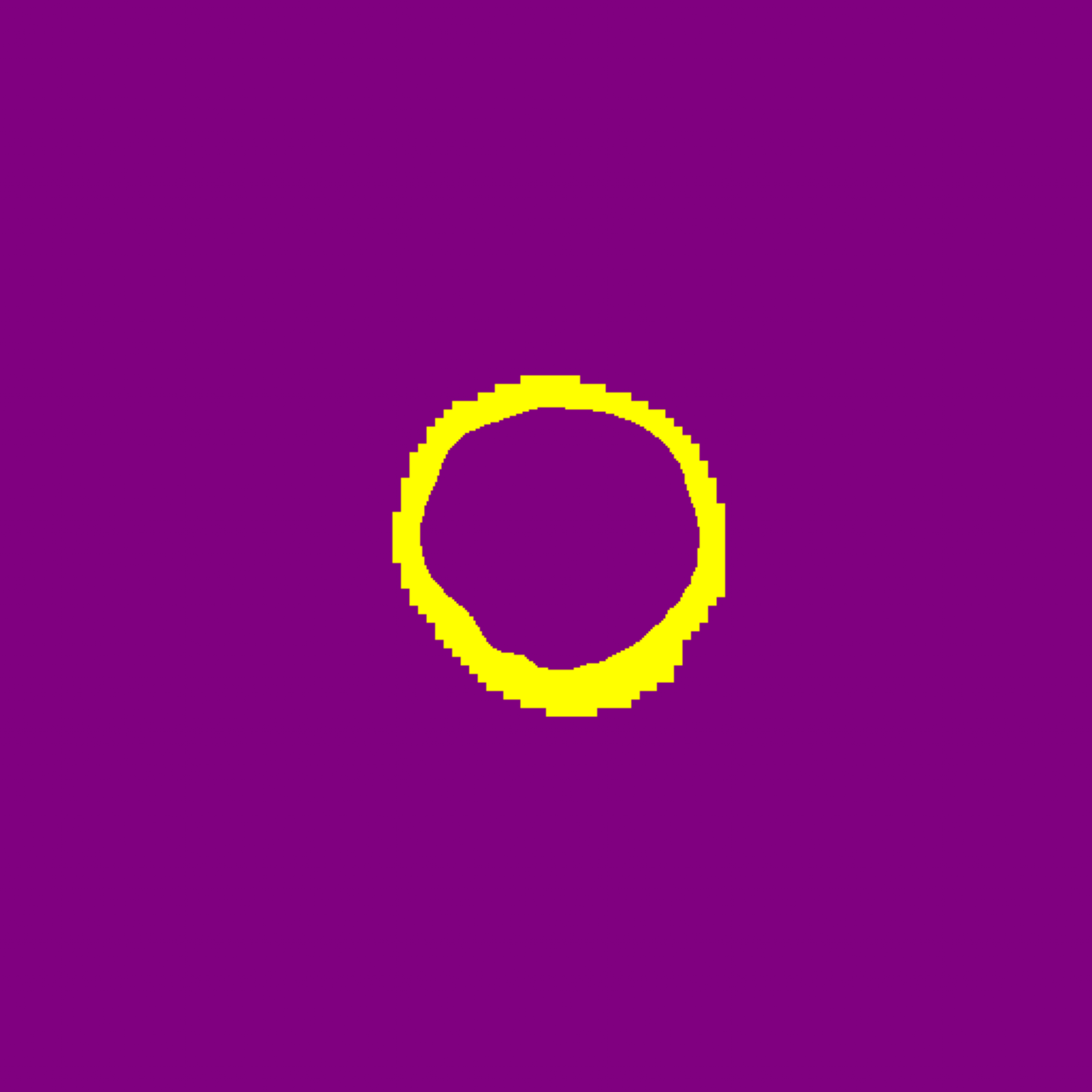} \\
        
        \raisebox{1.9\height}{\rotatebox[origin=c]{90}{\tiny N-91-L}} &
        \includegraphics[width=\linewidth]{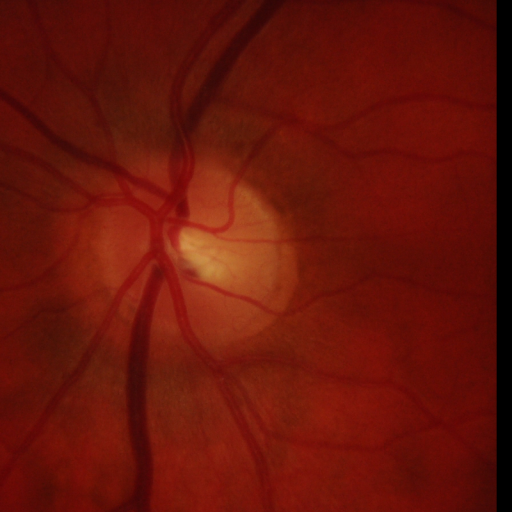} & 
        \includegraphics[width=\linewidth]{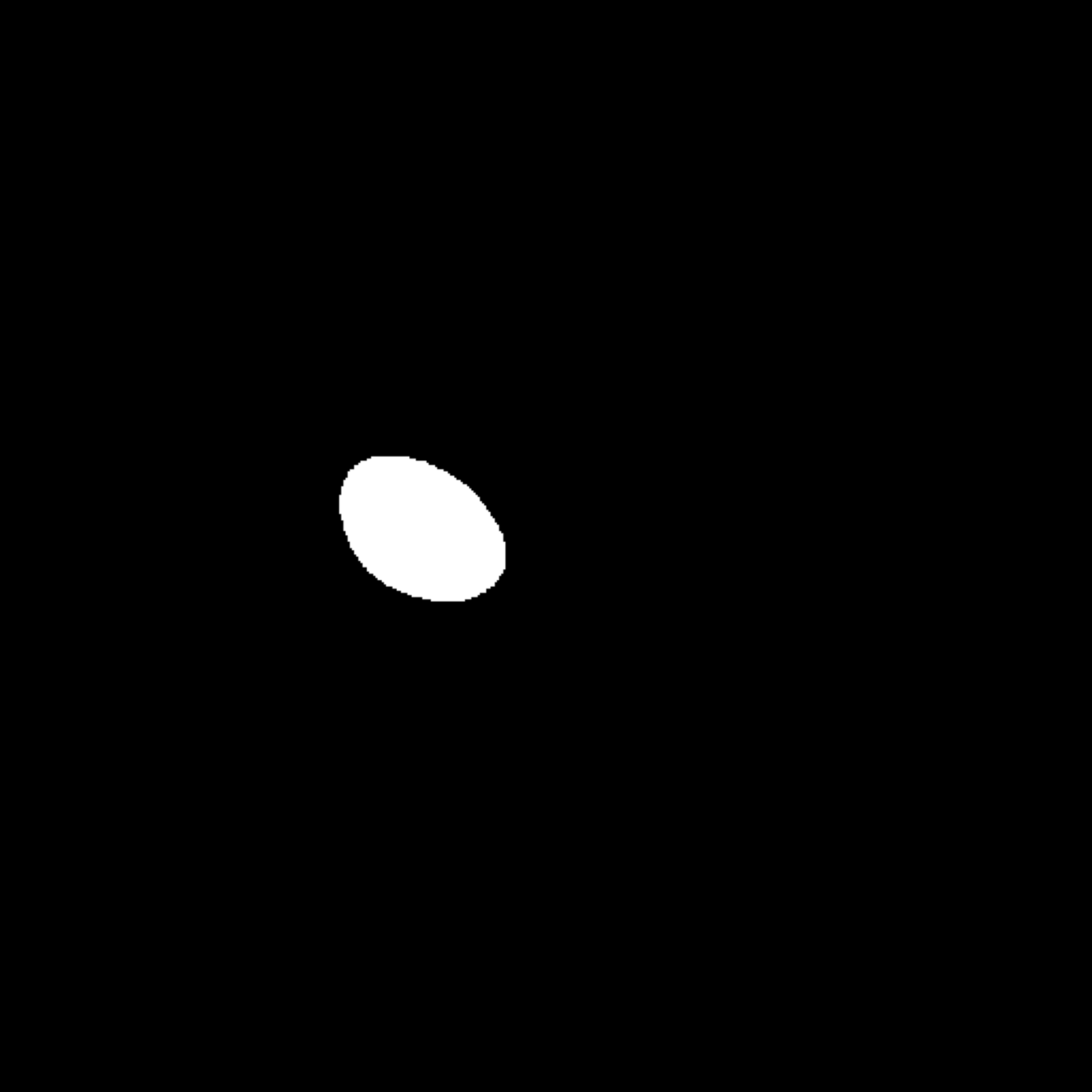} & 
        \includegraphics[width=\linewidth]{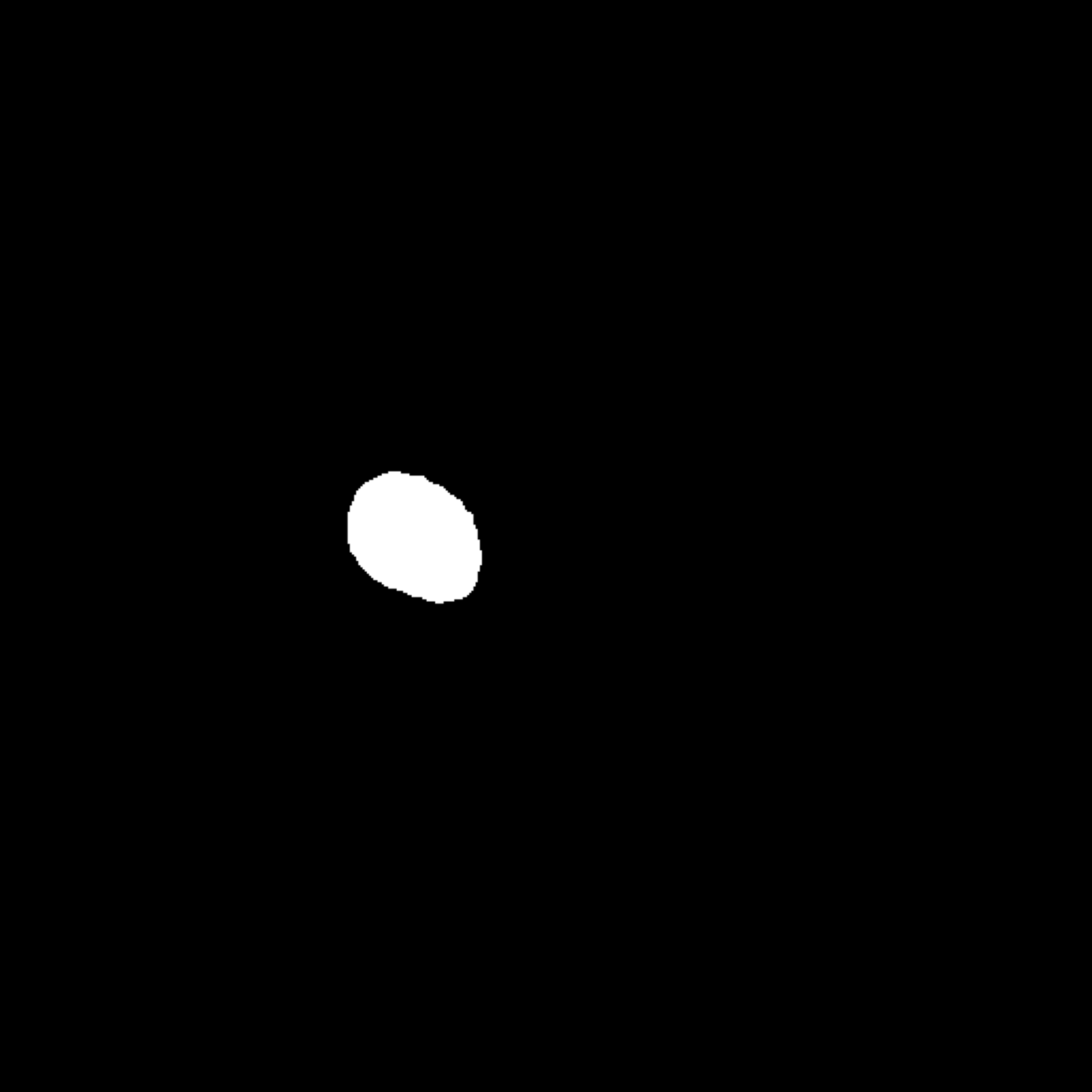} & 
        \includegraphics[width=\linewidth]{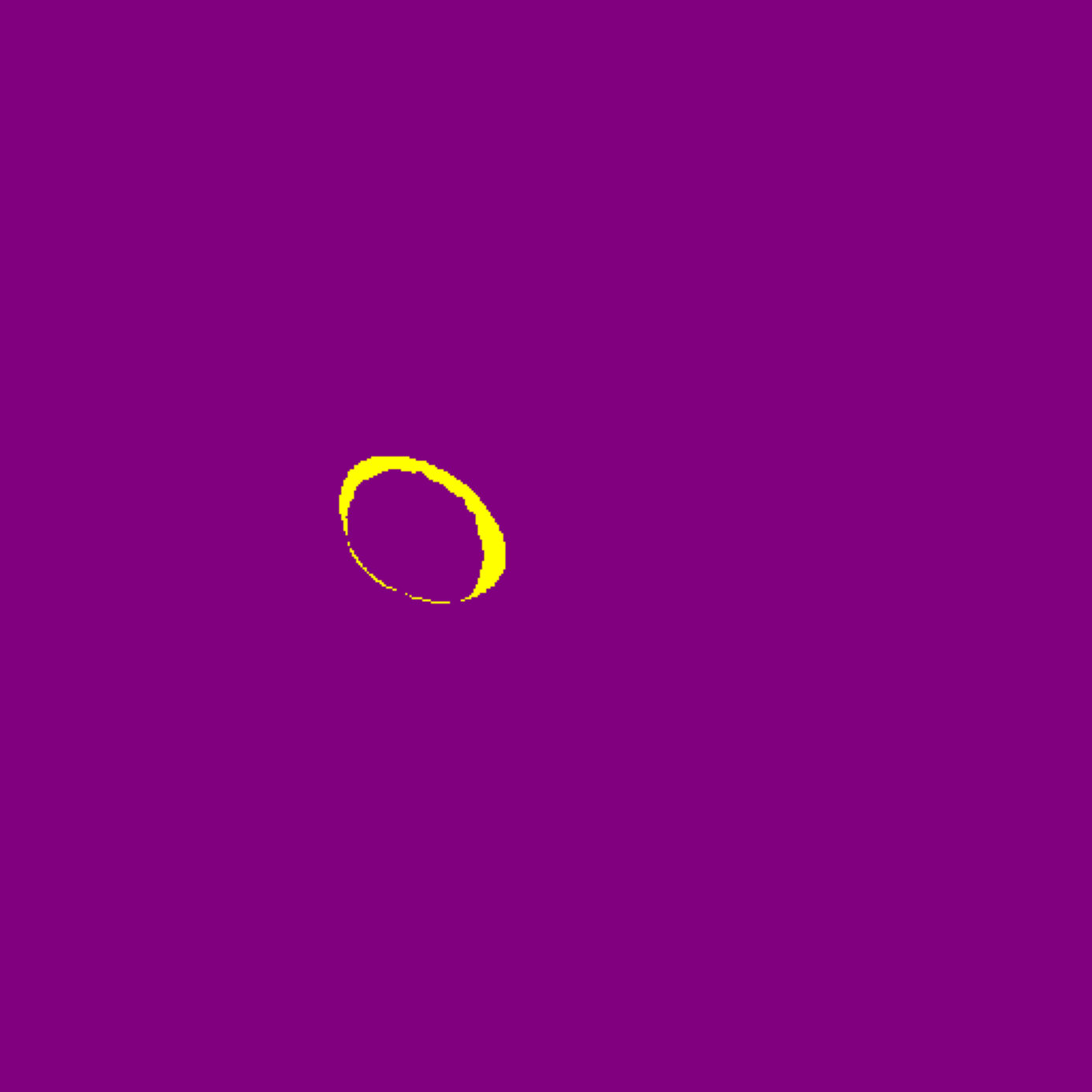} &
        \includegraphics[width=\linewidth]{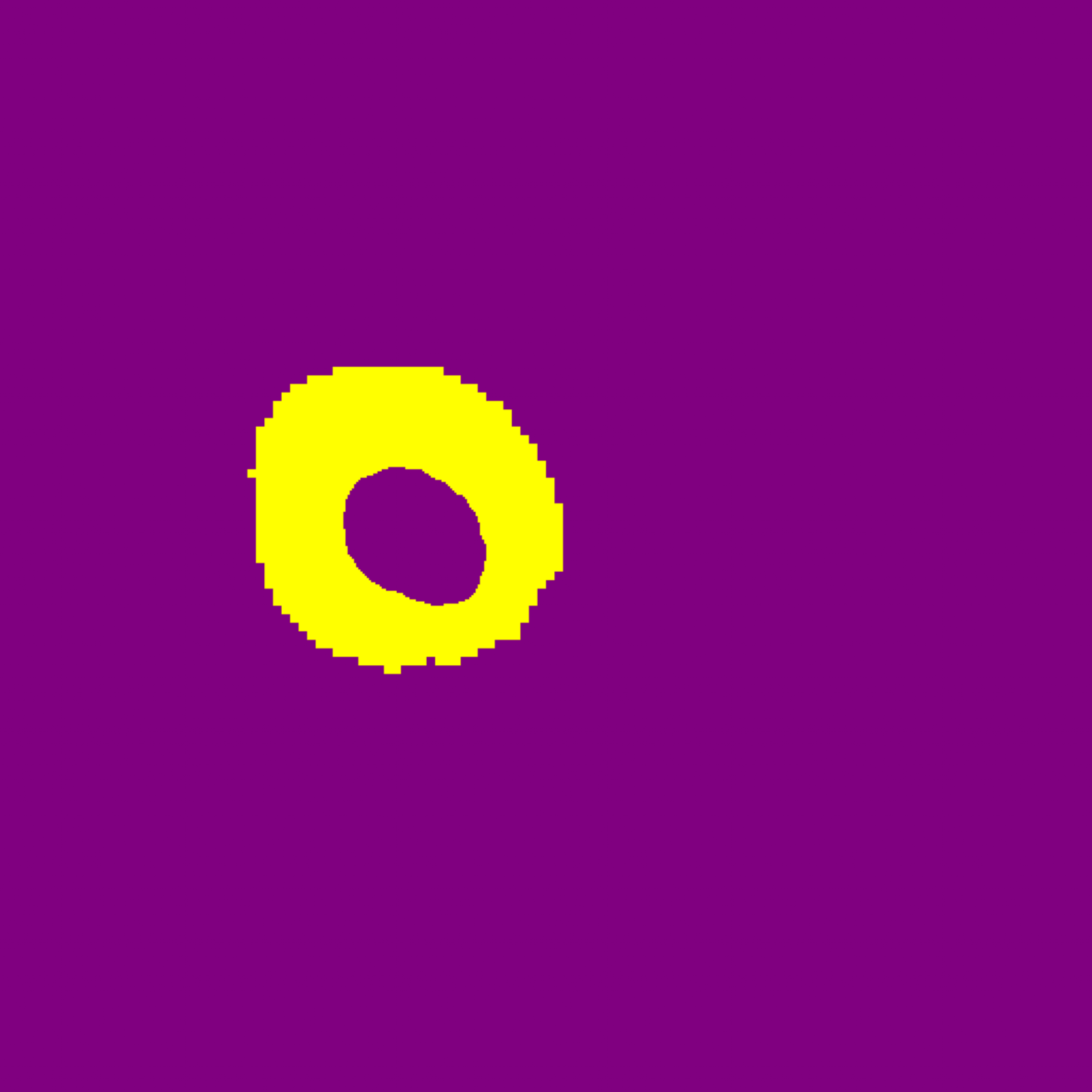} & 
        \includegraphics[width=\linewidth]{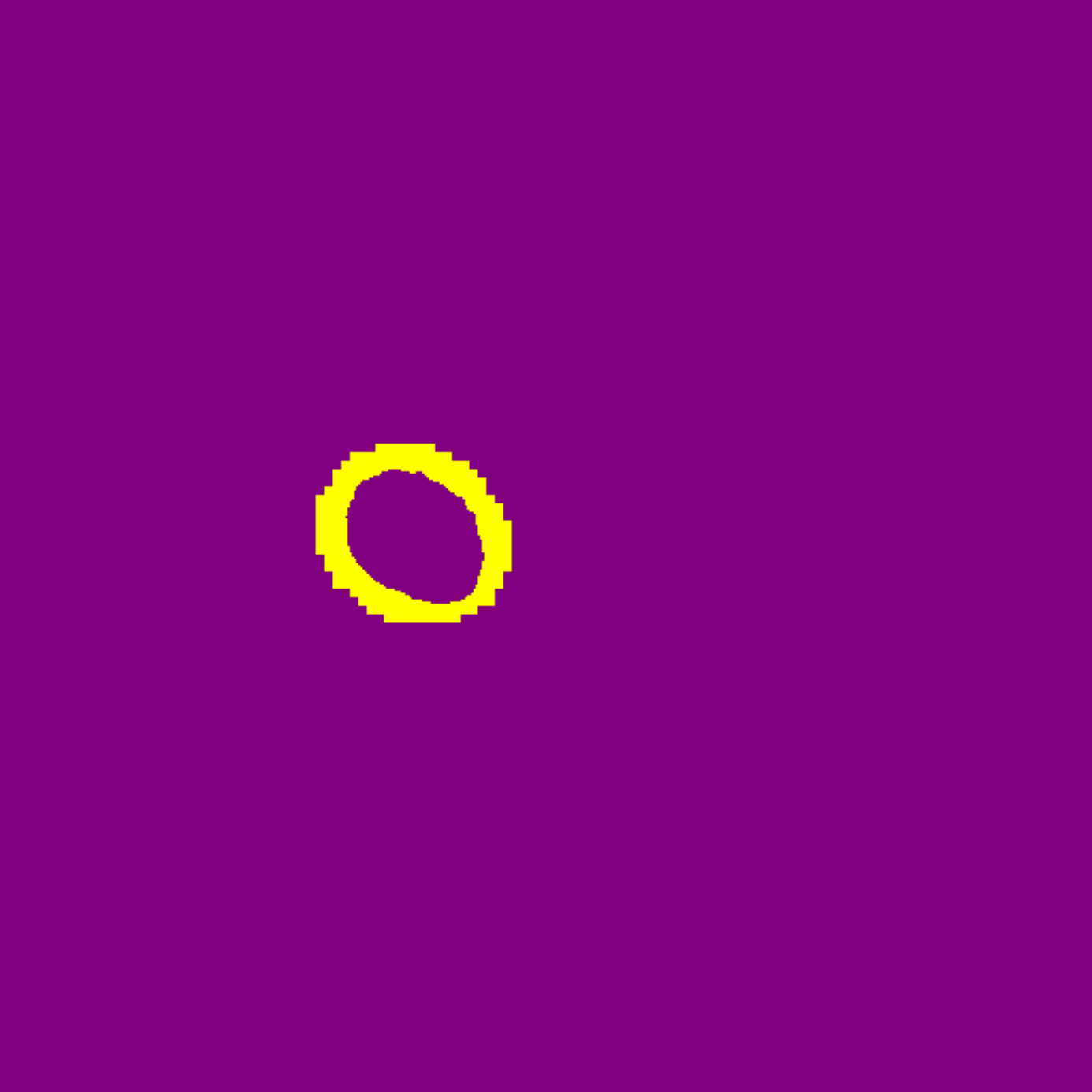} \\

        & \scriptsize Original Image  
        & \scriptsize Ground Truth  
        & \scriptsize Pseudo-Label 
        & \scriptsize Noisy Ground Truth
        & \scriptsize Baseline 
        & \scriptsize Refined Prototype Filtering (\textbf{Ours}) \\
    \end{tabular}
    }
    \caption{Qualitative comparison of denoise masks $m^{w_1}$ (Eq.~\ref{eq:denoise_mask}) for the \textit{cup} class. \textcolor{yellow}{Yellow} and \textcolor{purple}{Purple} indicate noisy and true pixel-level pseudo-labels, respectively. Column 4 shows noisy pseudo-label maps, Column 5 shows masks without filtering, and Column 6 shows masks using $m^{w_1}_{info\_region}$ and $m^{w_1}_{uncertainty}$ to retain informative, low-uncertainty regions.}
    \label{fig:3.1}
\end{figure*}

\subsection{Comparison with the State-of-the-Art}
We conducted extensive benchmarking and compared our method with existing UDA methods, including BEAL~\cite{BEAL} and AdvEnt~\cite{AdvEnt}, as well as state-of-the-art SFDA methods such as TENT~\cite{Tent}, DPL~\cite{DPL}, CPR~\cite{CPR}, CBMT~\cite{CBMT}, PLPB~\cite{PLPB}, and SBIF~\cite{yaacovi2025source} on fundus datasets. We also report results for the fully supervised \textit{Target Only} setting and the \textit{Source Only} baseline, which has no access to target images. BEAL~\cite{BEAL} mitigates uncertainty in soft boundary regions via adversarial learning. AdvEnt~\cite{AdvEnt} is a UDA method that enforces entropy consistency between source and target domains. TENT~\cite{Tent} leverages entropy minimization to adapt the model to the target distribution. DPL~\cite{DPL} introduces a denoising strategy to enhance self-training, thereby providing more discriminative and less noisy supervision. CPR~\cite{CPR} identifies context-inconsistent predictions and proposes a context-aware pseudo-label refinement mechanism to improve adaptation. PLPB~\cite{PLPB} designs a pseudo-boundary loss to exploit edge information from both domains, enabling more accurate predictions in boundary regions. Recently, SBIF~\cite{yaacovi2025source} leveraged the foundation model SAM~\cite{SAM} to evaluate pseudo-label quality and mitigate noisy supervision during target model training.

\begin{table}[!ht]
\centering
\caption{Ablation study of diffrent components in the framework (REFUGE $\rightarrow$ Drishti-GS / RIM-ONE-r3): Average Dice and ASSD. \textbf{Bold} texts highlight the best scores, \underline{Underlined} texts highlight the second-best scores.}
\begin{tabular}{l|cc|cc}
\toprule
\multirow{2}{*}{\textbf{Configuration}} & \multicolumn{2}{c|}{\textbf{Drishti-GS}} & \multicolumn{2}{c}{\textbf{RIM-ONE-r3}} \\
& Dice$\uparrow$ & ASD$\downarrow$& Dice$\uparrow$& ASD$\downarrow$\\
\midrule
Vanilla & 89.09 & 8.12 & 83.38 & 10.79 \\
+ RPF & 90.65 & 7.16 & 87.94 & 5.94 \\
+ EntropyFilt & 90.04 & 7.65 & 84.89 & 7.44 \\
\hspace{4mm} + RPF & 90.91 & 6.93 & 88.07 & 5.86 \\
+ UG-EMA & 90.25 & 7.50 & 85.43 & 8.74 \\
\hspace{4mm} + EntropyFilt & 90.37 & 7.65 & 85.68 & 8.65 \\
\hspace{4mm} + RPF & \underline{91.28} & \underline{6.57} & \underline{88.75} & \underline{5.44} \\
+ All (Full) & \textbf{91.64} & \textbf{6.28} & \textbf{89.21} & \textbf{5.18} \\
\bottomrule
\end{tabular}
\label{tab:components}
\end{table}

As shown in Table~\ref{tab:comparison_results}, our method achieves state-of-the-art performance on both Drishti-GS and RIM-ONE-r3, effectively narrowing the gap with the \textit{Target Only} model. The improvements are most pronounced for challenging classes like the optic cup, demonstrating the effectiveness of our pseudo-label refinement and boundary-focused design. On Drishti-GS, the gains for optic disc segmentation are smaller due to its larger and more consistent structure, but our approach excels when the foreground is small or dominated by background pixels, highlighting its strength in refining fine boundaries.

We also evaluate our method under an open-domain setting (Table~\ref{tab:open_avg}), where it maintains strong generalization and achieves state-of-the-art results. Compared to the second-best method, our model improves Dice by 2.02\% and reduces ASSD by 2.48 in the C+O setting, confirming that our pseudo-label refinement remains reliable even for unseen data.

Notably, our method outperforms BEAL, a vanilla UDA approach that requires access to source data. This can be attributed to the challenge UDA faces in learning invariant source–target features, whereas our method directly optimizes on target data for better adaptation.

The qualitative results in Figure~\ref{fig:qualitative_results} (two samples per target domain) demonstrate that our method delivers the best overall performance. While most models produce boundaries with residual artifacts or missing pixels, particularly around challenging edges, our model closely aligns with the ground truth, significantly reducing such errors. Moreover, in the open-domain setting, our approach successfully captures target boundaries and achieves predictions that nearly match the ground truth.

\subsection{Ablation Studies}

\textbf{Components.} Table~\ref{tab:components} evaluates the contribution of each module. The baseline teacher-student model benefits most from adding \textit{RPF}, confirming that robust pseudo-label filtering is critical for reducing error accumulation. Combining \textit{RPF} with \textit{UG-EMA} further improves performance by selectively updating the teacher only when the student provides reliable signals. Adding \textit{EntropyFilt} yields the best results, as it focuses learning on ambiguous regions rather than already confident predictions.

\textbf{RPF in Pseudo-Label Denoising.} Figure~\ref{fig:3.1} illustrates how \textit{RPF} improves noisy pseudo-label detection compared to the baseline, particularly in dense or uncertain regions. By refining the boundaries, \textit{RPF} produces cleaner supervision that aligns more closely with the noisy ground truth, effectively highlighting the misclassified pixels in the pseudo-label when compared to the ground truth. In contrast, the baseline either incorrectly filters out the entire surrounding region of the cup or fails to identify the noisy pixels, which leads to missing important details around the foreground and accumulating errors. This behavior can be attributed to the ambiguity of boundary regions in the feature space: unlike certain background regions with distinct characteristics, boundary areas lack clear separability. Consequently, during clustering, their proximity to the foreground prototype rather than the background prototype often results in misclassification.

\textbf{UG-EMA Metrics.} As shown in Table~\ref{tab:metrics}, entropy weighted by the inverted Gaussian function outperforms both raw entropy and loss-based metrics. This weighting prioritizes boundary pixels, which are more informative, while downplaying overly confident central regions.

\begin{figure}[!ht]
    \centering
    \includegraphics[width=0.8\linewidth]{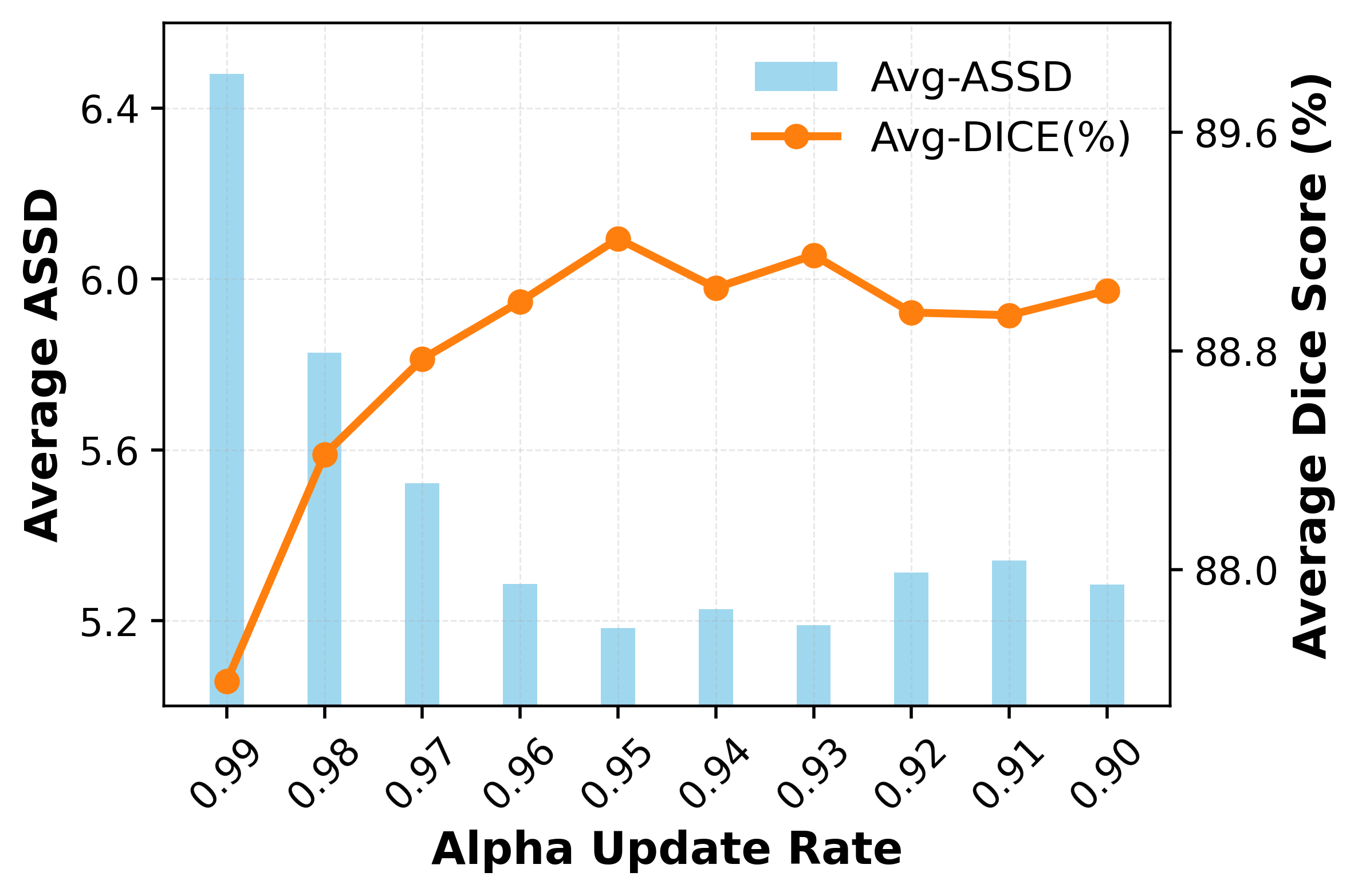}
    \caption{UG-EMA update rate $\alpha$ on REFUGE $\rightarrow$ RIM-ONE}
    \label{fig:alpha_metrics}
\end{figure}

\textbf{UG-EMA Hyper-parameter.} We also investigate the impact of different values for the $\alpha$ update rate in UG-EMA, as shown in Figure~\ref{fig:alpha_metrics}, which illustrates the model's performance on the RIM-ONE dataset. Interestingly, our model's performance increases gradually and becomes more stable as the $\alpha$ update rate decreases, despite the teacher model being updated more aggressively. This indicates that our mechanism, UG-EMA, effectively controls the teacher model, ensuring that the quality of the pseudo-labels improves throughout the adaptation stage. In contrast, a high $\alpha$ rate prevents the teacher from learning any knowledge from the student model, resulting in the pseudo-label's quality not improving significantly.

\begin{figure}[!ht]
    \centering
    \includegraphics[width=0.8\linewidth]{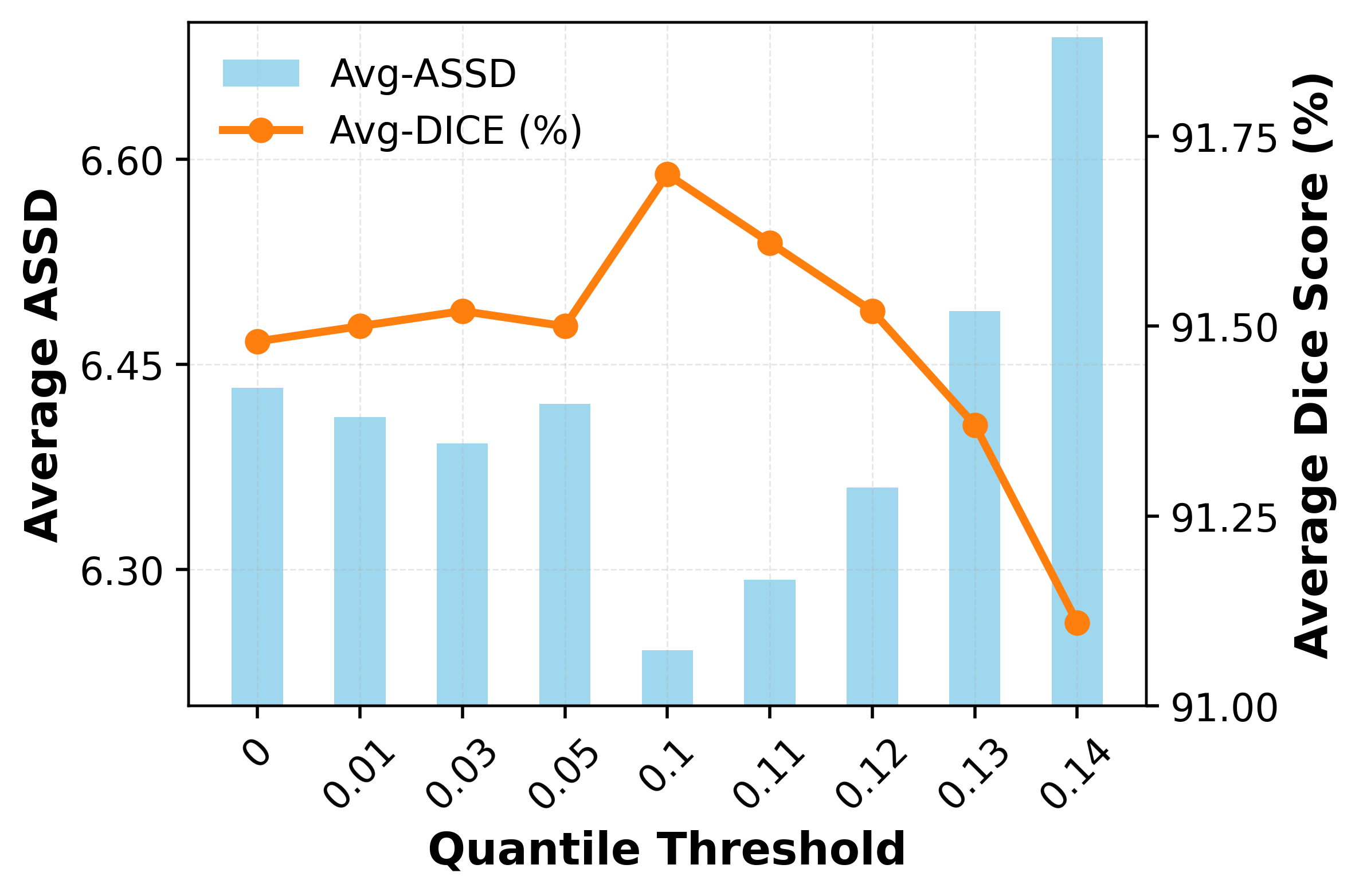}
    \caption{The $\beta$ quantile theshold to filter out high confidence prediction on REFUGE $\rightarrow$ Dritish}
    \label{fig:quantile_threshold}
\end{figure}

\textbf{Quantile threshold.} 
We further investigate the effect of the $\beta$ quantile threshold by starting from $0$ (no filtering) and gradually increasing the threshold step by step. Experiments are conducted on the Drishti-GS dataset, as shown in Figure~\ref{fig:quantile_threshold}, where our model achieves the best performance at $\beta = 0.1$. This suggests that selecting an appropriate filtering threshold allows the model to perform effectively. However, as the threshold increases further, the model begins to filter out more informative pixels, leading to significant performance degradation. In contrast, when using a very low threshold, from no filtering up to around $0.1$, it has only a mild effect, likely because the selected pixels still include values close to $0$ and $1$, preventing the model from focusing exclusively on high-entropy regions.

\textbf{Hyper-parameter Scale}
\begin{table*}[ht]
\centering
\begin{tabular}{c|ccccccc}
\hline
\textit{s} & 0.15 & 0.2 & 0.25 & 0.3 & 0.35 & 0.4 & 0.45 \\
\hline
Average Dice & 89.07 & 89.00 & 89.06 & 88.97 & 89.08 & 89.05 & 89.03 \\
\hline
\end{tabular}
\caption{Dice score with different scale values on RIM-ONE-r3 dataset. Our method is robust to the hyperparameter setting.}
\label{tab:dice_scale}
\end{table*}

Table~\ref{tab:dice_scale} shows that the scale parameter has minimal impact ($<0.12\%$ Dice variation), confirming that the inverted Gaussian weighting is robust and retains its focus on boundary refinement across different settings.

\begin{figure}[!ht]
    \centering
    \includegraphics[width=0.7\linewidth]{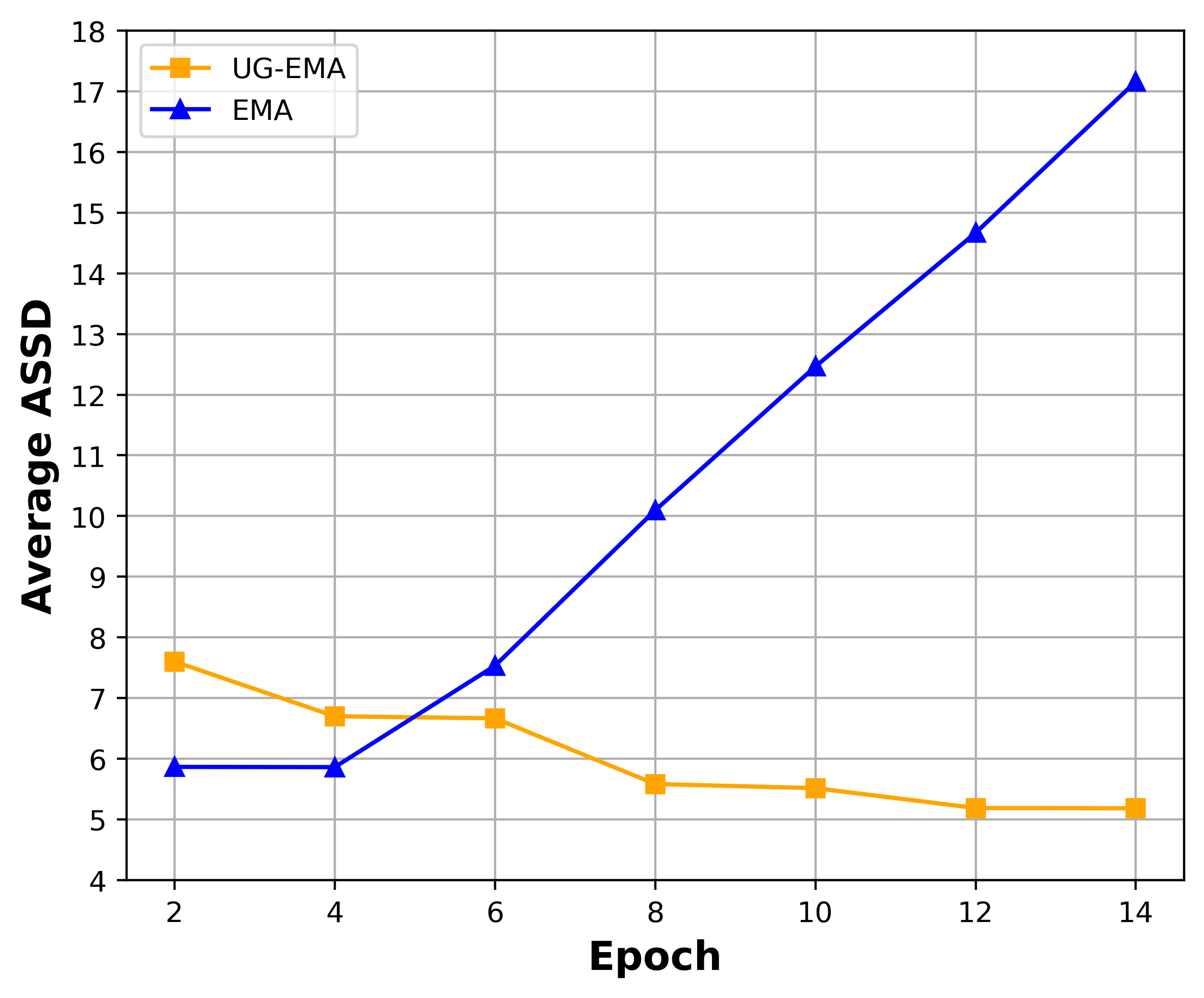}
    \caption{Student training progress of REFUGE $\rightarrow$ RIM-ONE (ASSD).}
    \label{fig:sub_assd}
\end{figure}

\begin{figure}[!ht]
    \centering
    \includegraphics[width=.7\linewidth]{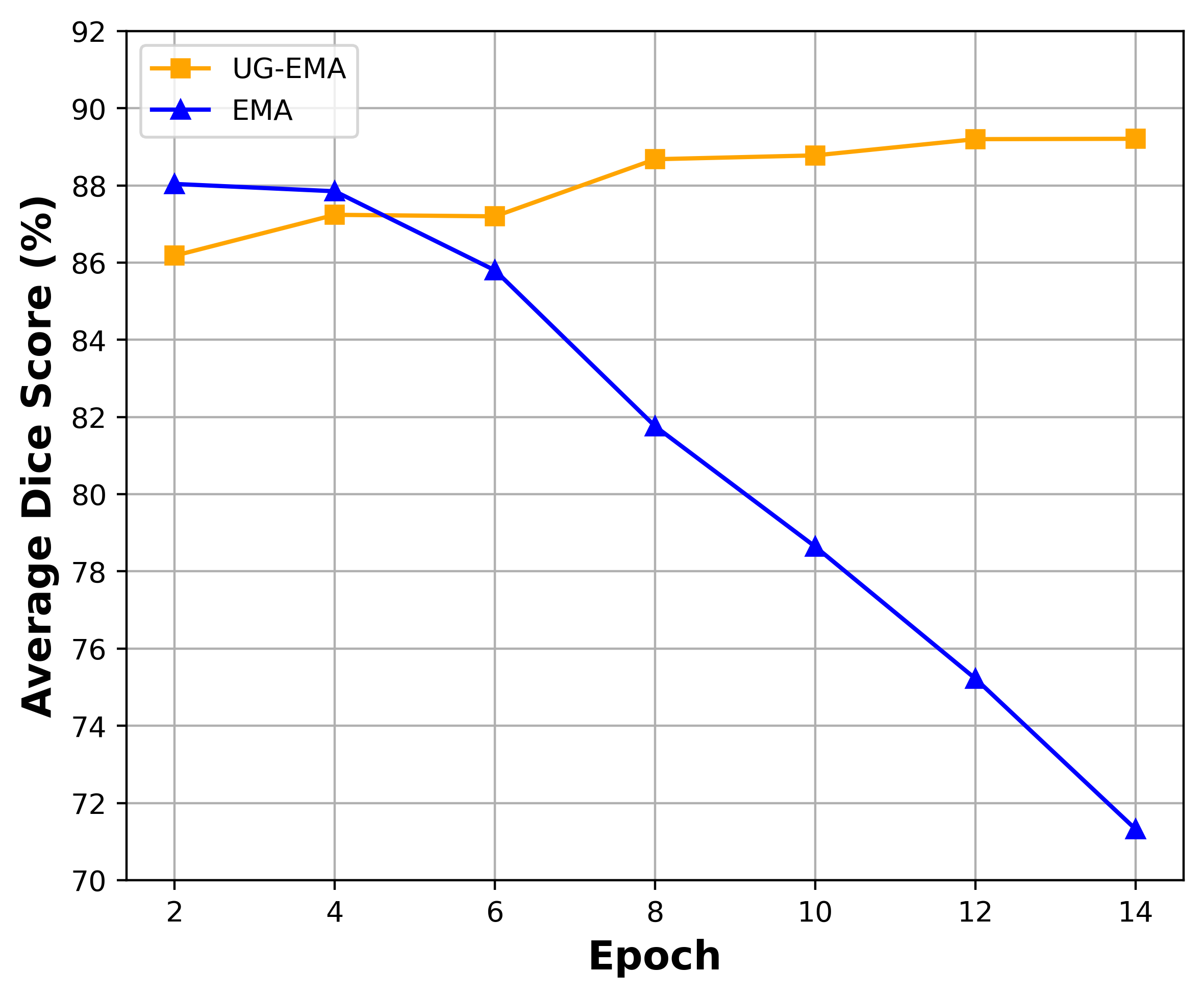}
    \caption{Student training progress of REFUGE $\rightarrow$ RIM-ONE (Dice).}
    \label{fig:sub_dice}
\end{figure}

\textbf{Training progress.} We visualize the adaptation process using ASSD in Figure~\ref{fig:sub_assd} and Dice in Figure~\ref{fig:sub_dice}. The overall trend shows that the UG-EMA method is more stable and maintains better performance toward the end of the adaptation stage. In contrast, with the same update rate, EMA suffers from error accumulation because it updates all versions of the student model, including erroneous predictions. This degrades the quality of the pseudo-labels produced by the teacher over the course of the adaptation process.

%% file: sec/conclusion.tex
\section{Conclusion}
We introduced a novel algorithm for source-free domain adaptive medical image segmentation. Our approach, called UP2D, leverages a student-teacher semi-supervised learning architecture where the teacher denoises pseudo-labels through a newly designed Refined Prototype Filtering mechanism that prioritizes informative, low-uncertainty regions, while we also proposed an Uncertainty-Guided EMA strategy to prevent error accumulation. This strategy selectively updates the teacher based on reliable student predictions. To improve boundary precision and generalization, we employed a quantile-based entropy filtering technique to focus learning on ambiguous regions. Extensive experiments on multiple medical benchmarks demonstrate that our method achieves SoTA performance(s), particularly on challenging classes dominated by background pixels.

\section{Acknowledgment}
This work was supported in part by the U.S. NIH under Grants NCI U01-CA268808 and NHLBI R01-HL171376. We thank AI VIETNAM for supporting us GPUs to conduct the experiments.